\documentclass[twocolumn]{aastex631} 
\usepackage{subfigure}
\usepackage{amsmath}
\usepackage{multirow}
\usepackage{tablefootnote}
\usepackage{microtype}

\newcommand{\FT}{}

\newcommand{\apjrv}{}

\begin{document}

\title{Nonthermal Pressures: Key to Energy Balance and Structure Formation Near SgrA* in the Milky Way}

\correspondingauthor{Farideh Mazoochi}
\email{famazoochi@ipm.ir}

\author[0000-0003-3390-4893]{Farideh Mazoochi}
\affiliation{Institute for Research in Fundamental Sciences (IPM), School of Astronomy, Tehran, Iran}

\author[0000-0002-0377-0970]{Fatemeh S. Tabatabaei}
\affiliation{Institute for Research in Fundamental Sciences (IPM), School of Astronomy, Tehran, Iran}
\affiliation{Max-Planck-Institut f\"ur Radioastronomie, Auf dem H\"ugel 69, D-53121 Bonn, Germany}

\author[0000-0003-0410-4504]{Ashley~T.~Barnes}
\affiliation{European Southern Observatory (ESO), Karl-Schwarzschild-Stra{\ss}e 2, 85748 Garching, Germany}

\author[0000-0001-8064-6394]{Laura Colzi}
\affiliation{Centro de Astrobiología (CAB), CSIC-INTA, Carretera de Ajalvir km 4, Torrejón de Ardoz, 28850 Madrid, Spain}

\author[0000-0002-8586-6721]{Pablo Garc\'ia}
\affiliation{Chinese Academy of Sciences South America Center for Astronomy, National Astronomical Observatories, CAS, Beijing 100101, China}
\affiliation{Instituto de Astronom\'ia, Universidad Cat\'olica del Norte, Av. Angamos 0610, Antofagasta, Chile}

\author[0000-0002-7495-4005]{Christian Henkel}
\affiliation{Max-Planck-Institut f\"ur Radioastronomie, Auf dem H\"ugel 69, D-53121 Bonn, Germany}

\author[0000-0002-8455-0805]{Yue Hu}
\affiliation{Institute for Advanced Study, 1 Einstein Drive, Princeton, NJ 08540, USA}

\author[0000-0001-6353-0170]{Steven N. \textsc{Longmore}}
\affiliation{Astrophysics Research Institute, Liverpool John Moores University, IC2, Liverpool Science Park, 146 Brownlow Hill, Liverpool L3 5RF, UK}
\affiliation{Cosmic Origins Of Life (COOL) Research DAO, https://coolresearch.io}

\author[0000-0001-9281-2919]{Sergio Mart\'in}
\affiliation{European Southern Observatory, Alonso de C´ordova 3107, Vitacura 763 0355, Santiago, Chile}
\affiliation{Joint ALMA Observatory, Alonso de C´ordova 3107, Vitacura 763 0355, Santiago, Chile}

\author[0000-0002-3078-9482]{\'Alvaro S\'anchez-Monge}
\affiliation{Institut de Ci\`encies de l'Espai (ICE), CSIC, Campus UAB, Carrer de Can Magrans s/n, E-08193, Bellaterra (Barcelona), Spain}
\affiliation{Institut d'Estudis Espacials de Catalunya (IEEC), E-08860, Castelldefels (Barcelona), Spain}

\author[0000-0002-2887-5859]{Víctor M.Rivilla}
\affiliation{Centro de Astrobiología (CAB), CSIC-INTA, Carretera de Ajalvir km 4, Torrejón de Ardoz, 28850 Madrid, Spain}

\author[0000-0002-1730-8832]{Anika Schmiedeke}
\affiliation{Green Bank Observatory, 155 Observatory Road, Green Bank, WV 24944, USA}

\author[0000-0001-8224-1956]{Juergen Ott}
\affiliation{National Radio Astronomy Observatory, P.O. Box O, 1003 Lopezville Road, Socorro, NM 87801, USA}

\author[0000-0001-7330-8856]{Daniel~L.~Walker}
\affiliation{UK ALMA Regional Centre Node, Jodrell Bank Centre for Astrophysics, Oxford Road, The University of Manchester, Manchester M13 9PL, UK}

\author[0000-0002-9279-4041]{Q. Daniel \textsc{Wang}}
\affiliation{Department of Astronomy, University of Massachusetts, Amherst, MA 01003, USA}

\author[0000-0001-5933-2147]{Gwenllian M. Williams}
\affiliation{Department of Physics, Aberystwyth University, Ceredigion, Cymru, SY23 3BZ, UK}

\author[0000-0002-8389-6695]{Suinan Zhang}
\affiliation{Shanghai Astronomical Observatory, Chinese Academy of Sciences, 80 Nandan Road, Shanghai 200030, People’s Republic of China}

%



\begin{abstract}

The circumnuclear region of the Galactic Center offers a unique laboratory to study energy balance and structure formation around Sgr\,A$^{\star}$. This work investigates thermal and nonthermal processes within 7\,pc {\FT distance from Sgr\,A$^{\star}$}. Using MeerKAT 1.3\,GHz radio continuum data and ALMA H40$\alpha$ radio recombination line emission from the ACES survey, we separate free-free and synchrotron components at $\sim$0.2\,pc resolution. {\FT With a thermal fraction of $\simeq$13\%, the 1.3\,GHz emission shows tight correlations with the Herschel PACS infrared data. The correlation between the equipartition magnetic field and molecular gas traced by JCMT $^{12}$CO (J=3$\rightarrow$2) observations reveals a balance between the magnetic field, cosmic rays, and molecular gas pressures south of the circumnuclear disk on $\sim$0.7\,pc scales. Unlike the magnetic field and ionized gas, the molecular gas density declines in the cavity (R$\leq 2$\,pc) toward the center, likely due to feedback from Sgr\,A$^{\star}$.} 
We find that nonthermal pressure from turbulent gas nearly balances magnetic and cosmic ray pressures and exceeds thermal pressure by two orders of magnitude. {\FT The medium surrounding Sgr\,A$^{\star}$ is filled} by a low-$\beta$ (thermal-to-magnetic energy), supersonic plasma, with an Alfvén Mach {\FT number $\simeq 4$ (assuming equipartition)}. Analysis of the mass-to-magnetic flux ratio suggests that the circumnuclear region is mostly subcritical and, therefore, the magnetic field can help stabilize gas clouds against gravitational collapse. 

\end{abstract}



\section{Introduction} \label{sec.intro}


 The Galactic Center (GC) at the heart of the Milky Way serves as a prominent region for star formation, and is the galactic nucleus where resolved studies of the interstellar medium (ISM) are possible with most common instruments \citep{yusef2013alma,yusef2015signatures,henshaw2022star}.
The star formation activities at the GC have been discussed in various studies \citep[e.g.][]{morris1993massive,kruijssen2014controls,barnes2017star,henshaw2022star,king2024chimps2,2024A&A...690A.121C}. Although this central region is rich in molecular gas that fuels star formation, the star formation efficiency (SFE) in the GC is lower than anticipated \citep{longmore2013variations}.
A high gas threshold density ($n_{\rm th}=10^7 {\rm cm}^{-3}$) for star formation suggests that the slow evolution of the gas toward collapse is limiting the star formation rate (SFR) due to a combined effect of several factors in this region \citep[e.g.][]{langmore,kruijssen2014controls}. The role of magnetic fields and turbulence has been studied as the most prominent factors controlling the star formation activity in molecular clouds of the central molecular zone  \citep[CMZ, e.g.,][]{pillai2015magnetic,federrath2016link,lu2024magnetic}.  The question of whether turbulence or magnetic fields have a greater impact on the formation of clouds and stars remains unanswered. This study aims to address this question by investigating the physical conditions of the ISM within the inner 7\,pc distance from the super-massive black hole (SMBH) of the Milky Way, Sgr\,A$^{\star}$, through a detailed study of thermal and nonthermal processes.

The circumnuclear region in the vicinity of the SMBH is a fascinating area that deserves a specific study. Within this region, the circumnuclear disk (CND) is the largest and closest molecular gas structure to Sgr\,$\rm A^{\star}$ that follows a quasi-Keplerian rotation \citep{liu2012milky,martin2012surviving}. The disk extends from an inner radius of approximately 1.5–2\,pc and {\apjrv likely} does not exceed 7\,pc in extent \citep{genzel1985neutral,jackson1993neutral,christopher2005hcn}. A strong magnetic field as well as supersonic motions are characteristic properties of the CND \citep{guerra2023strength,akshaya2024magnetic}.  The neutral gas in the disk has densities of $\sim 10^5$\,cm$^{-3}$, and the total hydrogen mass is $\rm \sim10^4\,M_{\odot}$ with a temperature of a few 100\,K \citep{genzel1985neutral,mezger1996galactic}. 

A mini-spiral structure exists in the inner $\sim$2\,pc of this central region (Sgr\,A West), with $\rm A_v$ varying from 20 to 50 mag with a median value of 31.1 mag \citep{scoville2003hubble}. Due to tidal shear, star formation near the SMBH is difficult to initiate. Nevertheless, within the central parsec, in the vicinity of Sgr\,A$^{\star}$, a massive stellar cluster is observed \citep{schodel2009nuclear,yusef2015signatures}. Investigating the physical condition of the ISM helps us to unravel mysteries in this central region.


In this study, we aim to answer some fundamental questions: 
What role do the thermal and nonthermal processes play in shaping the ISM structures in the GC? How are different ISM components interconnected, and what is the impact of the SMBH on these correlations, and up to which galactocentric radius are these correlations detectable?
The first step to answer our questions is separating the two components, synchrotron and free-free, of the radio continuum (RC) emission. The RC emission traces the nonthermal processes via its synchrotron component, which is radiated by relativistic cosmic ray electrons (CREs) spiraling around the magnetic fields in the ISM. The power-law spectrum of this radiation is described by a variable nonthermal spectral index $\alpha_n$  ($S_{\nu}\propto \nu^{-\alpha_n}$) in an optically thin condition, and depends on the energy power-law index of CREs. 
The free-free component of the RC emission arises from a thermally ionized medium. In an optically thin condition ($\tau_{\nu}<<1$), the spectral energy distribution of the free-free emission follows a flat power-law relation with a spectral index of approximately $-0.1$ ($S_{\nu}\propto\nu^{-0.1}$).  In the frequency range of $\rm \nu<~10\,GHz$, the nonthermal component dominates the RC emission from normal star-forming galaxies \citep{condon1992radio,tabatabaei2017radio}.
By separating the thermal and nonthermal components of the RC emission, we can gain a deeper understanding of the ISM energy and pressure balance within galaxies \citep{ghasemi2022evolution,hassani2022role}. This separation allows us to investigate the magnetic field strength in different regions and explore the relationship between the magnetic field and the efficiency of star formation \citep{tabatabaei2018discovery}. 

The comprehensive analysis of the RC emission within the GC reveals a combination of the thermal and nonthermal emission \citep[e.g.,][]{law2008green}. Various techniques, such as extrapolating high-frequency emission and fitting spectral energy distributions with an assumed and fixed spectral index, are used to separate the thermal and nonthermal components in different regions of the GC \citep{law2008green,yusef2012interacting,meng2019physical}. However, the nonthermal spectral index $\alpha_n$ can vary in different environments due to different energy loss mechanisms of CREs \citep{longair2010high}. Considering a constant $\alpha_n$ in resolved studies can lead to an underestimation or an overestimation of the thermal fraction. This study uses the thermal radio template (TRT) method \citep{tabatabaei2007high,tabatabaei2013detailed} in the circumnuclear region as one of the most accurate and reliable methods. Unlike the classical methods, the TRT method takes into account the variability of $\alpha_n$, leading to more accurate results. Based on the TRT method, the recombination line emission is used to trace the thermal emission. Subtracting the thermal emission from the observed RC, the nonthermal RC component is obtained.

 The data used in this survey are introduced in Section~\ref{sec.data}. Section~\ref{sec.separation} describes the separation of the thermal and nonthermal intensities. In Section~\ref{sec:analysis}, the radio and IR correlations are investigated in the circumnuclear region. It also details the calculation of the magnetic field strength from the nonthermal component, along with the calculation of the thermal ionized gas density and the molecular gas column density. Section~\ref{sec.dis} discusses the radial distribution and relationships of the obtained properties, the energy balance, and mass-to-magnetic flux ratio in this central region. Finally, Section~\ref{sec.conclusion} presents the conclusions.

\section{Data}\label{sec.data}
  In this study, we used a combination of the radio and IR observations of the circumnuclear region, which are outlined in Table~\ref{table:1}.

\begin{table*}[ht]
\caption{\FT Summary of data used in this study}            
\label{table:1}      
\centering                          
\begin{tabular}{c c c c c c c}        
\hline\hline                 
\multirow{2}{*}{Data} &Frequency/& Angular& \multirow{2}{*}{RMS noise}& Calibration&\multirow{2}{*}{Telescope} & \multirow{2}{*}{Reference}\\
&Wavelength&Resolution&&noise&\\
\hline 
\hline
   RC &1.3\,GHz& $4^{''}\times 4^{''}$& 4.75$\times10^{-5}$[Jy\,beam$^{-1}$]&5\%& MeerKAT& \cite{heywood20221}\\      
   
   
   H40$\alpha$ &99.0\,GHz& $1.4^{''}\times 1.2^{''}$& 1.25$\times10^{-1}$[Jy\,beam$^{-1}$\,km\,s$^{-1}$]&5\%& ALMA&Hsieh et al. in prep\\

   CO J=1$\rightarrow$0&115.3\,GHz& $15^{''}\times 15^{''}$&  86.2 [K\,km\,s$^{-1}$]&6.4\%& NRO &\cite{tokuyama2019high}\\

   CO J=3$\rightarrow$2&345.8\,GHz& $15^{''}\times 15^{''}$& 29.7 [K\,km\,s$^{-1}$]&6\%& JCMT &\cite{2020MNRAS.498.5936E}\\
   
   MIR &21.3\,$\mu m$& $18.3^{''}\times 18.3^{''}$& 1.42$\times10^{-6}$ [W\,m$^{-2}$\,sr$^{-1}$]&6\%& SPIRIT III&\cite{price2001midcourse}\\
   
   FIR &70\,$\mu$m&$6^{''}\times6^{''}$&0.14 [Jy \,pixel$^{-1}$]&5\%&Herschel&\cite{molinari2010hi}\\
   
   FIR &160\,$\mu$m& $12^{''}\times12^{''}$&0.35 [Jy\,pixel$^{-1}$]&5\%&Herschel&\cite{molinari2010hi}\\
   
\hline                                   
\end{tabular}
\end{table*}

The RC map of our study area was observed with the South African MeerKAT radio telescope using its L-band receiver \citep[frequency range of 856-1712 MHz][]{heywood20221}. This total intensity map is a 16-pointing hexagonal mosaic covering 6.5 square degrees of the central region with an angular resolution of $4^{''}$ at the central frequency of 1.3\,GHz. 
The calibration uncertainty of this instrument is 5\% \citep{knowles2022meerkat}.

To trace the free-free emission, a millimeter recombination line observed by the Atacama Large Millimeter/submillimeter Array (ALMA) is used. This line is the H40$\alpha$ radio recombination line (RRL) map from the ALMA CMZ Exploration Survey (ACES) at a frequency of 99.0\,GHz and {\apjrv an} angular resolution $\sim1.5^{''}$. ACES is a cycle 8 Band 3 Large program on ALMA (2021.1.00172.L,209 PI: S. Longmore) and observed the CMZ in the frequency range of 86-101\,GHz. This survey covers the CMZ spanning approximately $-0.6^{\circ}<l<+0.9^{\circ}$ and $-0.3^{\circ}<b<+0.2^{\circ}$ (Longmore et al. in prep). In this work, we used the data cube from the combination of ALMA 12m+7m+TP arrays and created the zero-moment integrated map intensity in the radial velocity range of -150\,km\,s$^{-1}$ to +150\,km\,s$^{-1}$ as \cite{tsuboi2018alma} did for H42$\alpha$  (also with ALMA data) to prevent contamination from other lines. We adopt the ALMA calibration uncertainty of 5\%.

The molecular gas of the GC is studied through $^{12}$CO (J=1$\rightarrow$0) \citep{tokuyama2019high} observations with the Nobeyama Radio Observatory (NRO) 45 m telescope and the $^{12}$CO (J=3$\rightarrow$2) \citep{2020MNRAS.498.5936E} data taken with of the James Clerk Maxwell Telescope (JCMT). The J=1$\rightarrow$0 line data of $^{12}$CO ($\nu_{\rm rest}$= 115.3\,GHz) covers the region within $-0.8^\circ \leq l \leq 1.4^\circ$ and $|b| \leq 0.35^\circ$. The obtained data has a $15^{''}$ full width half-maximum (FWHM) beam size and spectral resolution of 0.67\,km\,s$^{-1}$ integrated over velocities between $\rm V_{LSR}$ = -220 and +220\,km s$^{-1}$. The CO J=3$\rightarrow$2 transition at 345.8\,GHz is part of the CO Heterodyne Inner Milky Way Plane Survey 2 (CHIMPS2). This data was observed by the Heterodyne Array Receiver Program on the JCMT covering $-3^{\circ}\leq l\leq 5^{\circ}$ and $|b|\leq 0.5^{\circ}$ with a spectral and angular resolution of  1\,km\,s$^{-1}$ and $15^{''}$ over velocities of $\rm |V_{LSR}| \leq 300\,km \,s^{-1}$. The calibration uncertainties are 6\% for the NRO and 6.4\% for the JCMT. For this study, the $^{12}$CO (J=1$\rightarrow$0) data were only used to investigate the molecular gas kinematics. For cross-correlating the molecular gas emission with the data at other wavelengths, we used the JCMT $^{12}$CO (J=3$\rightarrow$2) data because of its better Nyquist sampling. On the other hand, the $^{12}$CO (J=3$\rightarrow$2) line may be much less affected by diffuse foreground gas than the $^{12}$CO (J=1$\rightarrow$0) line and is therefore also preferable.

The Midcourse Space Experiment (MSX) survey at 21.3\,$\mu$m in the Mid-infrared (MIR) wavelength is used in this study {\apjrv as the tracer of warm dust}. The infrared telescope, known as SPIRIT III, is mounted at the center of the spacecraft. The wavelength mentioned is associated with band E, with a calibration uncertainty of approximately 6\%. The angular resolution of this survey is $\sim 18.3^{''}$ and it covers the entire Galactic Plane within $|b|\leq 5^{\circ}$ \citep{price2001midcourse}.

In this study, we used the 70 and 160\,$\mu$m emissions of the Herschel infrared Galactic Plane (Hi-GAL) survey. This survey covers the Galactic Plane at longitudes of $|l|\leq 60^{\circ}$ and latitudes of $|b|\leq 1^{\circ}$, spanning five wavebands between 70 and 500\,$\mu$m \citep{molinari2010hi}. The 70 and 160\,$\mu$m emissions at resolutions of 6$^{''}$ and 12$^{''}$, respectively, are tracers of cold dust in the far-infrared (FIR) wavelength and are observed with the PACS photometric camera. The calibration uncertainties of this survey are about 5\% at 70\,$\mu$m and 160\,$\mu$m \citep{molinari2016hi}.

\begin{figure*}[!ht]
	\centering
	\includegraphics[width=1\textwidth]{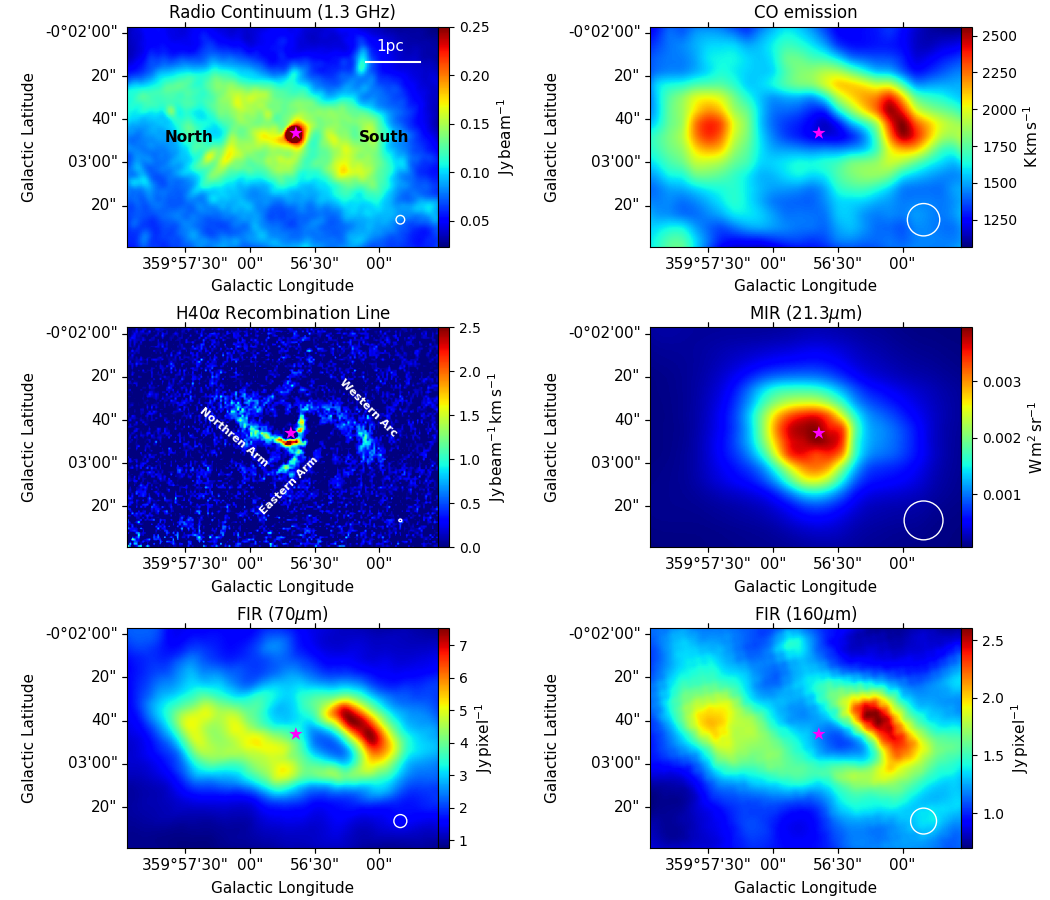}
	\caption{\FT The circumnuclear region (R$<$7\,pc) mapped by different observations: {\it Left, from top to bottom-} MeerKAT RC emission at 1.3\,GHz, ALMA H40$\alpha$ RRL integrated intensity at 99.0\,GHz, and Herschel FIR emissions at 70\,$\mu$m. {\it Right, from top to bottom-} JCMT $^{12}$CO (J=3$\rightarrow$2) integrated intensity at 345.8\,GHz, MSX MIR emission at 21.3\,$\mu$m, and  Herschel FIR emissions at 160\,$\mu$m. Circles in the lower right corners indicate the resolutions or beam sizes (see Table~\ref{table:1}). Stars indicate the position of Sgr\,A$^{\star}$.  The northern and southern parts of this region are indicated in the RC map. The names of the components within the mini-spiral structure are labeled on the RRL map.}
	\label{fig:data}
\end{figure*}

{\apjrv In} Section~\ref{sec.separation}, we obtain maps of the thermal and nonthermal components at $4^{''}$ angular resolution ($\simeq$ 0.2\,pc). 
Then, to compare these maps with those tracing dust and gas in the ISM, 
all data were convolved to the coarsest resolution of 18.3$^{\prime\prime}$ ($\simeq$0.7\,pc) of the MSX data 
and projected to the same geometry and pixel size. Figure~\ref{fig:data} shows different maps of the inner GC at R$<$7\,pc from radio to MIR at their original resolutions. 

Uncertainties in the observed intensities ($\sigma$) include both the statistical (root mean square noise, $\sigma_{rms}$ of the observed maps) as well as the systematic (flux calibration uncertainty of the instruments, $\sigma_{cal}$) errors. The total uncertainty for each data point is calculated using $\sigma = \sqrt{(\sigma_{cal}\times F_{\nu})^2+\sigma_{rms}^2}$, where $F_{\nu}$ represents the flux density at frequency $\nu$.  To determine the errors in the obtained parameters in this study, the $\sigma$ uncertainties are propagated through the calculations presented in the paper. Uncertainties due to assumptions are discussed separately.

\section{Mapping Thermal \& Nonthermal Emission}\label{sec.separation} 

{\apjrv In} this section, we map the thermal and nonthermal components of the RC emission at 1.3\,GHz 
using a thermal radio tracer. As the region of our study is highly extincted by dust, optical tracers such as H$\alpha$ recombination lines cannot be used. 
Unlike H$\alpha$, RRLs do not suffer from extinction and therefore provide a more robust tracer for the free-free emission in dusty regions \citep{scoville2003hubble}. 

The free-free continuum emission and the recombination line radiations are both linearly proportional to the ionizing UV photons if the ionized medium is optically thick to Lyman photons \citep{Osterbrock,Rubin}. As shown by \cite{Dickinson}, the ionized gas in the Galaxy is optically thick to ionizing Lyman photons \citep[i.e., case B,][]{Osterbrock} even for diffuse features at intermediate and high
Galactic latitudes ($1\,<\tau_{\rm Ly\alpha}<$\,30). 

In the TRT method, to measure thermal emission under local thermal equilibrium (LTE), we first need to calculate the emission measure (EM) using the following relation \citep{rohlfs2013tools}, 

\begin{equation} \label{eq:1}
\rm
    \frac{\rm T_L \Delta v}{\rm K\,km\,s^{-1}} = 5.76\times10^{2} \left(\frac{T_e}{K}\right)^{-1.5} \left(\frac{EM}{cm^{-6}\,pc}\right)\left(\frac{\nu_L}{GHz}\right)^{-1}, 
\end{equation}
where $\rm T_L \Delta v$  is the RRL integrated line intensity over the velocity width of the line for each pixel (expressed in units of $\rm K\,km\,s^{-1}$), \(\nu_L\) represents the frequency of the RRL, and $\rm T_e$ denotes the electron temperature. We note that Equation~(\ref{eq:1}) is valid if $\rm T_{L} = T_e \tau_L$, i.e. that the medium is optically thin ($\tau_L<<1$) to the RRL emission\footnote{We approximately checked the optical depth for H40$\alpha$ in the region of this study to confirm that $\tau_{\nu}<1$.}.

On the other hand, the optical depth of the continuum emission at the frequency (\(\nu_c\)) can be obtained from the EM using the following formula \citep{oster1961emission}:

\begin{equation} \label{eq:ta_c}
\rm
\tau_c = 8.23\times10^{-2}\times\left(\frac{T_e}{K}\right)^{-1.35} \left(\frac{EM}{cm^{-6}pc}\right)\left(\frac{\nu_c}{GHz}\right)^{-2.1}. 
\end{equation}

The brightness temperature of the thermal radio continuum emission, denoted as $\rm T_c$, is calculated using the radiative transfer equation: 

\begin{equation}
\rm T_c^{th} = T_e (1 - e^{-\tau_c}).   
\end{equation}

In the Rayleigh-Jeans limit, we convert the brightness temperature to flux density measured in Jy\,beam$^{-1}$ in order to obtain the thermal emission. Then, we subtract this thermal emission from the total continuum emission at the continuum frequency to determine the nonthermal flux density ($S_{\nu_c}^{\text{nt}} = S_{\nu_c}^{\text{RC}} - S_{\nu_c}^{\text{th}}$). In this study, we used the MeerKAT 1.3\,GHz map for the total continuum emission.

It should be noted that, at low frequencies, the emission of RRLs can be stimulated by nonthermal effects and generated by masing at high atomic levels; therefore, they are not suitable as pure thermal templates \citep{gordon1990observational,scoville2013submillimeter}. For example, \cite{meng2019physical} studied and compared RRLs at frequencies below 10\,GHz in the Sgr\,B2 region and found that these RRLs are likely excited under non-LTE conditions. According to \cite{scoville2013submillimeter}, the maser amplification is negligible for millimeter RRLs at energy levels of n=20-50, and the separation results derived from these lines are more reliable than those of higher levels. This fact highlights the importance of selecting an appropriate RRL as a thermal tracer to avoid overestimating the thermal emission. 

\begin{figure}[!ht]
    \centering
    \includegraphics[width=\columnwidth]{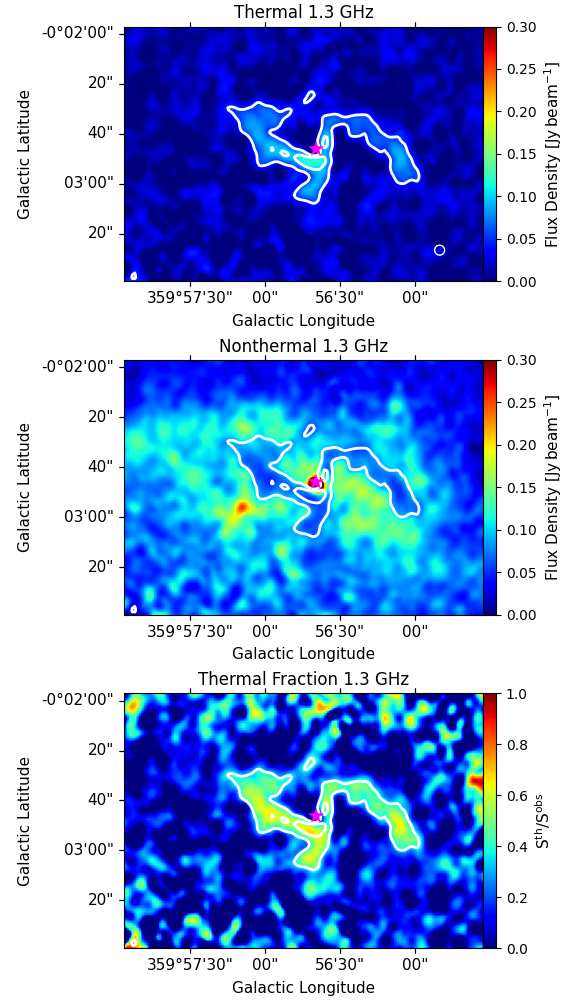}
    \caption{Maps of the thermal free-free emission (top), nonthermal synchrotron emission (middle),  and {\FT thermal fraction}  (bottom) at 1.3\,GHz in the circumnuclear region. The beam size of $4\arcsec \times 4\arcsec$ 
    is shown in the lower right corner of the first map. {\FT Contours in all maps show the thermal intensity at levels of 0.05, 0.1 Jy\,beam$^{-1}$. Stars indicate the position of Sgr\,A$^{\star}$.}}
    \label{pic:separat}
\end{figure}

We use the H40$\alpha$ recombination line  ($\nu_L$=99\,GHz)  as our thermal tracer in Equation~(\ref{eq:1}) to measure the EM. As shown by previous studies, the adopted $\rm T_e$ values range between 5000\,K and 7000\,K in this  region \citep{roberts1993multiconfiguration,simpson1997infrared,lang2001vla,scoville2003hubble,hsieh2018magnetic}. Hence, we use the average value of $\rm T_e=6000$\,K. 
{\apjrv The} resulting maps of the thermal and nonthermal emission are shown in Figure~\ref{pic:separat}, exhibiting different distributions. 
At the same intensity levels, the nonthermal emission has a larger extension. {\apjrv Part of this nonthermal emission can emanate from the  Sgr\,A East supernova remnant \citep{maeda2002chandra}, which is located very close (in projection) to the CND. The shock waves from Sgr\,A East can accelerate particles to relativistic speeds, producing synchrotron radiation. Another famous source of the nonthermal radiation in this region is Sgr\,$\rm A^{\star}$, which is a nonthermal compact source. The central $4^{"}$ beam, which contains the Sgr\,$\rm A^{\star}$, has a nonthermal flux of 0.44$\pm$0.05\,Jy, which contributes to the total nonthermal emission in our studied region by 0.8\,\%. {\FT Although it is unresolved at the resolutions of this study ($4^{"}$ and $18^{"}$), its feedback can influence its surroundings (See Section~\ref{sec:radial}). }

On the other hand,} the thermal fraction ($f_{\nu_c}^{\rm th}={S_{\nu_c}^{\rm th}}/{S_{\nu_c}^{\rm RC}}$) at 1.3\,GHz is highest at the mini-spiral structure ($f_{\rm 1.3\,GHz}^{\rm th}\simeq 40-60\%$, Figure~\ref{pic:separat}-bottom). The mean thermal fraction is 13.1$\pm$0.2\% over the entire region of this study. The error for the mean of $f_{\rm 1.3\,GHz}^{\rm th}$ is the standard deviation divided by the square root of the sample size. We note that a change in $\rm T_e$ between 5000 and 7000\,K results in an uncertainty of less than 20\% in $f_{\rm 1.3\,GHz}^{\rm th}$. {\apjrv Previous} studies reported the thermal fractions by assuming a constant nonthermal spectral index in a more extended area of the GC and at different frequencies than in this work. 
For example, \cite{law2008green} reported a thermal fraction of about 19\% at 5\,GHz and 28\% at 8\,GHz for individual sources within $4^{\circ}\times10^{\circ}$ of the GC. 

\begin{figure}[!ht]
	\centering
	\includegraphics[width=1\columnwidth]{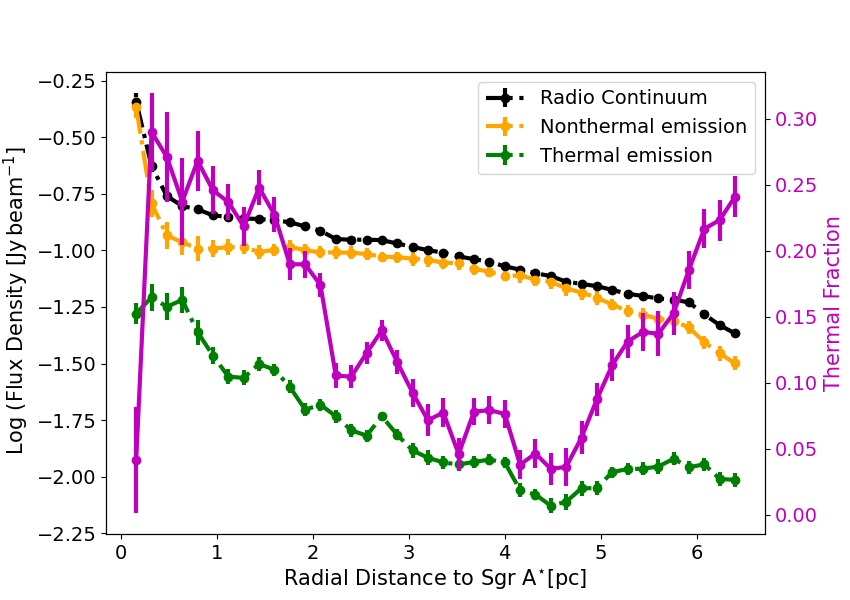}
	\caption{\FT Radial profiles of the RC emission (mean intensities in rings of 4\arcsec~width vs radial distance to Sgr\,A$^{\star}$) and its thermal and nonthermal components at 1.3\,GHz. Also shown is the radial profile of the thermal fraction ({\it magenta}). Error bars represent the standard deviation divided by the square root of the sample size per ring.}

	\label{fig:th-nth}
\end{figure}

{\apjrv Figure~\ref{fig:th-nth}} shows the $4\arcsec$ ring mean intensities of the observed, thermal, and nonthermal emission at 1.3\,GHz against galactocentric radius from Sgr\,A$^{\star}$. This radial distance (R) is calculated assuming an inclination of $72^{\circ}$ (see Section~\ref{sec.mf}) and a distance of 8200\,pc to the GC \citep{leung2023measurement}. The nonthermal synchrotron emission is brighter than the thermal emission everywhere, especially at the position of Sgr\,A$^{\star}$ where $f_{\rm th}<5\%$. In the CND (1.5$<$R$<$7\,pc), the mean thermal fraction is 11.8$\pm$0.2\%. {\apjrv Starting from $r\simeq 0.5$\,pc, the nonthermal emission shows a sudden increase toward the center, indicating the influence of Sgr\,A$^{\star}$ feeding its surroundings by injecting CREs (see also Section~\ref{sec:radial}).}   

\section{Analysis}\label{sec:analysis}
{\apjrv In} this section, {\apjrv we} first investigate any correlation between the IR bands and different components of the RC emission. Then,  we derive the equipartition magnetic field strength using the nonthermal synchrotron emission. The resulting thermal map offers insights into the volume density of {\apjrv thermally} ionized gas. Additionally, we estimate the molecular gas column density using the JCMT CO observations.

\subsection{Radio-IR Correlation}\label{sec: radio-IR}

\begin{figure*}[!ht]
	\centering
	\includegraphics[width=1\textwidth]{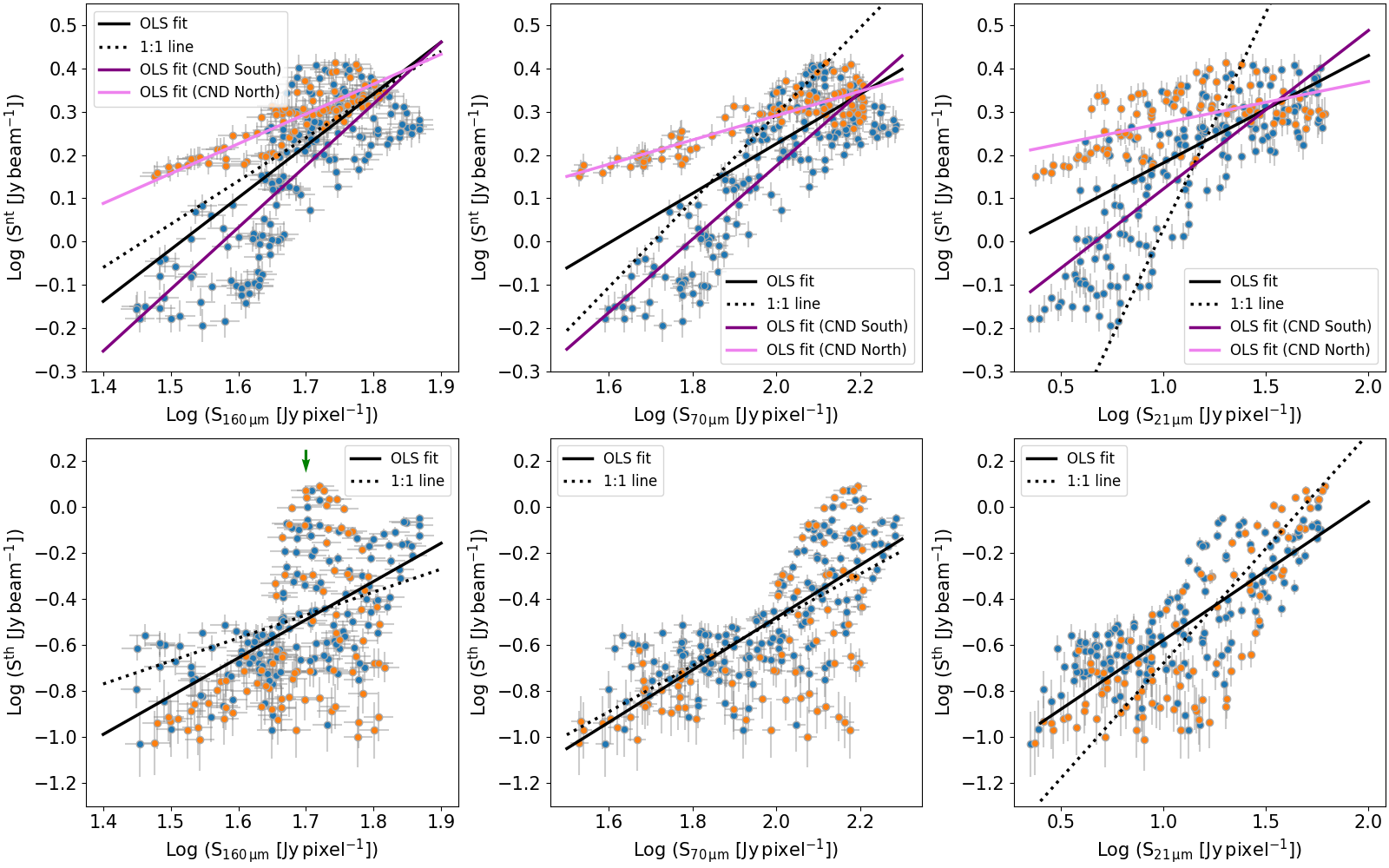}
	\caption{Relationships between the nonthermal ($\rm S^{nt}$, {\it top}) and thermal ($\rm S^{th}$, {\it bottom}) intensities at 1.3\,GHz and the 160, 70, and 21\,$\mu$m IR emissions in the circumnuclear region. Solid black lines indicate the OLS fits, and dotted lines represent the 1:1 line. Orange and blue data points represent the northern and southern parts. Pink and purple lines show the OLS fits to the north and south, respectively. The green arrow referred to the vertical distribution.}
	\label{fig:FIR-nth}
\end{figure*}

The tight correlation between nonthermal radio and thermal IR emission, first inferred by a linear relation between integrated luminosities of galaxies,  is linked to massive star formation \citep[e.g.,][]{Helou,dejong,Gavazzi}. The radio--IR correlation was later found on $\lesssim$kpc scales inside galaxies as well \citep[e.g.,][]{huges,murphy_06,dumas} indicating an interplay between massive star formation and the magnetized/relativistic ISM \citep{taba_2012,tabatabaei2013detailed,taba_13}.  These resolved studies showed that an almost linear correlation\footnote{A correlation between two quantities with a linear dependency.} can hold even in weakly star-forming regions, indicating that a balance between CREs, magnetic fields, and gas pressures is a prerequisite for this linearity, regardless of the level of star formation \citep{taba_13}. As such, studying this correlation on resolved scales provides, in general, a useful tool for inferring the propagation length of CREs and the mixing/coupling of magnetic field and gas \citep{taba2013,nasirzadeh2024radio}. Hence, we investigate any equipartition condition in this region by studying the nonthermal radio--IR correlation (see Section~\ref{sec.mf}). Comparing the RC 1.4\,GHz with the IR 60\,$\mu$m integrated fluxes, \cite{Crock11a} found that the RC was relatively weak compared to that expected from the global radio-IR correlation in the inner 200\,pc of the Milky Way, linking it to winds/outflows. {\apjrv In Figure~\ref{fig:FIR-nth}, we} investigate this correlation locally (pixel-by-pixel) between the different RC components at 1.3\,GHz and the IR bands from 21$\mu$m to 160$\mu$m as tracers of warm and cold phases of the ISM, {\apjrv respectively.}

{\apjrv The} Pearson correlation coefficient between two images $S_{1}(x,y)$ and $S_{2}(x,y)$ of identical pixels $i$ is given by:
\begin{equation}
r_{p} = \dfrac{\sum_{i=0}^{\rm n} (S_{1i}-\langle S_{1}\rangle)(S_{2i}-\langle S_{2}\rangle)}{\sqrt{\sum_{i=0}^{\rm n}(S_{1i}-\langle S_{1}\rangle)^2\sum_{i=0}^{\rm n} (S_{2i}-\langle S_{2}\rangle) ^{2}}},
\end{equation}
with 
 \(S_{1i}\) and \(S_{2i}\) representing the intensity of the $i$-th pixel in images 1 and 2, respectively. The mean of the pixel intensity values in each image is denoted as $\langle S_1 \rangle = \frac{1}{n} \sum_{i=0}^n S_{1i}$ and $\langle S_2 \rangle = \frac{1}{n} \sum_{i=0}^n S_{2i}$, with \(n\)  the total number of pixels. For a perfect correlation (anti-correlation), $r_{p}=$1 ($r_{p}=$-1). If the two images are completely uncorrelated, then  $r_{p} = 0$. The statistical error in $r_{p}$ depends on the strength of the correlation and the number of pixels n, $\triangle r_{p}$ = $\sqrt{1 -r_{p}^{2}}/({\sqrt{n-2}})$. The resulting Pearson correlations {\apjrv coefficients ($r_p$) 
 are listed in Table~\ref{tab2}. For these correlations, the central beam, which contains Sgr\,$\rm A^{\star}$, is subtracted to prevent any possible contamination from this source.}
 
In general, as evidenced by the obtained correlation coefficients ($r_p$ listed in Table~\ref{tab2}), the nonthermal synchrotron emission is well correlated with different IR bands tracing different ISM phases.  Fitting the linear curves $Y=\,c+m\,X$ in log-log planes, we obtain the relationships between the radio emission vs the IR emission ($\rm S_2$ and $\rm S_1$ in Table~\ref{tab2}). A linear correlation, i.e., a correlation with a fitted power-law slope of 1, shows a similar distribution and a one-to-one variation of the two emissions in the medium. As mentioned above, finding an almost linear correlation can strengthen the possibility of using the equipartition assumption, which is needed to estimate the magnetic field strength in Section \ref{sec.mf}. The ordinary least squares (OLS) fits to the data show that the nonthermal synchrotron emission correlates with the 160$\mu$m emission almost linearly\footnote{The small super-linearity is linked to a more extended distribution of the synchrotron emission compared to that of the IR emission due to diffusion of CREs \citep{tabatabaei2022cloud}.} but sub-linearly ($m<$1) with the 21\,$\mu$m and 70\,$\mu$m emission (Table~\ref{tab2}). This indicates that the fine balance between the CREs, magnetic field, and gas is better achieved in the colder phase of the ISM traced by the 160\,$\mu$m emission than in the warmer phase by the 21 and 70\,$\mu$m emission. \cite{crocker2011wild} also found a similar result by studying the correlation between the 1.4\,GHz and 60\,$\mu$m emission. This can be understood because winds and outflows suggested by \cite{crocker2011wild} involve ionized gas and warm dust, while the hydrostatic pressure balance between cosmic rays and magnetic fields can be kept in other parts of the ISM, which are cooler than the outflow regions. 

As shown in Figure~\ref{fig:FIR-nth}, in addition to the general correlations, two distinct trends (blue and orange data points) are visible in the nonthermal--IR correlation plots, which correspond to the northern and southern parts of this central region. These two regions appear as bright structures in both the CO and FIR maps (Figure~\ref{fig:data}) and also exhibit strong nonthermal emission (Figure~\ref{pic:separat}). Fitting the nonthermal--IR correlation separately, we find that the northern part follows a flatter relation than the southern part (Table~\ref{tab2}), meaning that the lack of RC compared to IR is more severe in the north. This can be due to winds and outflows removing cosmic rays from the IR emitting clouds and matches with the previous finding of the presence of outflows in this northern CND region by \cite{zhao2016new}. 
 
The correlation between the thermal free-free emission and the IR emission is more or less as expected: A tighter correlation with warmer than with colder dust (see $r_p$ values listed in Table~\ref{tab2}). A linear thermal radio--IR correlation is generally expected if both emissions are tracing the SFR \citep[e.g.,][]{murphy_11}.  In the region of our study, such a linear relation is found with the 70$\mu$m emission, indicating that both emissions have a common origin in the ISM, i.e, star-forming regions (see more discussion in Section~\ref{sec:radial}). However, there is an excess of the 21$\mu$m emission with respect to the free-free emission ($m=0.57\pm0.02$). Contamination from Sgr\,A$^{\star}$ and/or the atmosphere of old carbon stars can be a possible reason for the excess of mid-IR radiation in this region. The fitted power-law slope of the thermal radio vs 160\,$\mu$m relation is steeper ($m = 1.70 \pm 0.14$), indicating an excess of diffuse ionized gas with respect to the cold dust in this region. This is evident from a population of points with a vertical distribution in Figure~\ref{fig:FIR-nth} (bottom-left), which belong to the ionized spiral structure except for the bright part of the western arc (Figure~\ref{fig:data}) \footnote{The ionized gas from the spiral structure except the bright western arc shows a correlation with the warmer dust emitting at 70$\mu$m.}. We note that, unlike the nonthermal--IR correlation,  the thermal--IR correlation does not show distinct trends in the northern and southern parts of this region (Figure~\ref{fig:FIR-nth}).


\begin{table}
	\centering
	\caption{Radio-IR correlations and best fitted relations in logarithmic scale in the circumnuclear region.}
	\begin{tabular}{ccccc}
		\hline\hline
            $\rm S_2$&$\rm S_1$&$m$\footnote{Slope of the fitted relations ${\rm log}(S_2)=\,c+m$ ${\rm log}(S_1)$  with $\rm S_1$ and $\rm S_2$ the IR and radio flux densities}&
            $r_p$\footnote{{\apjrv Pearson correlation coefficient}}&\\
		\hline
             $\rm S_{1.3\,GHz}^{\text{nt}}$&$\rm S_{160\,\mu m}$&1.20$\pm$0.09&0.68$\pm$0.03&\\
             &$\rm S_{70\,\mu m}$&0.57$\pm$0.04&0.69$\pm$0.03&\\
             &$\rm S_{21\,\mu m}$&0.25$\pm$0.02&0.64$\pm$0.04&\\
             \hline
             &\multicolumn{2}{c}{South}&\\
             &$\rm S_{160\,\mu m}$&1.35$\pm$0.08&0.79$\pm$0.03&\\
             &$\rm S_{70\,\mu m}$&0.85$\pm$0.04&0.86$\pm$0.02&\\
             &$\rm S_{21\,\mu m}$&0.37$\pm$0.02&0.80$\pm$0.03&\\
             \hline
             &\multicolumn{2}{c}{North}&\\
             &$\rm S_{160\,\mu m}$&0.69$\pm$0.03&0.85$\pm$0.03&\\
             &$\rm S_{70\,\mu m}$&0.28$\pm$0.02&0.86$\pm$0.03&\\
             &$\rm S_{21\,\mu m}$&0.10$\pm$0.01&0.54$\pm$0.08&\\
		\hline
            \hline
            $\rm S_{1.3\,GHz}^{\text{th}}$& $\rm S_{160\,\mu m}$& 1.70$\pm$0.14&0.51$\pm$0.04\\
 & $\rm S_{70\,\mu m}$& 1.07$\pm$0.06&0.70$\pm$0.03\\
 & $\rm S_{21\,\mu m}$& 0.57$\pm$0.02&0.77$\pm$0.02\\
 \hline
	\end{tabular}
	\label{tab2}
\end{table}

\subsection{Equipartition Magnetic Field}\label{sec.mf}
The nonthermal RC emission is related to synchrotron radiation and can be used to trace both the magnetic field strength $\rm B$ and the cosmic ray electron density $n_{\rm CR}$  \citep[$I_n \propto n_{\rm CR} \nu^{\alpha_n} B^{1-\alpha_n}$, ][]{pacholczyk1970radio}.
There are various methods to estimate magnetic field strengths within galaxies. A common method to estimate the magnetic field strength is using the synchrotron emission and assuming that the energy density of cosmic rays equals that of the magnetic field (B/CR equipartition), which is also considered in GC studies \citep[e.g., ][] {tsuboi1986prominent,larosa2004new,crocker2011wild,yusef2022statistical}. 

{\apjrv On} the other hand, this condition can be violated if CREs are cooled mainly by a mechanism different from the synchrotron cooling, such as bremsstrahlung energy loss in star-bursting regions with $\Sigma_{\rm SFR}\geq 100\,\rm M_{\odot}\,yr^{-1}\,kpc^{-2}$ \citep{yoast2016equipartition} or due to propagation of CREs on small scales \citep{seta2019revisiting}. However, the propagation length scale of these particles depends on the uniformity of the magnetic field in both streaming \citep{CRE1&Kulsrud, CRE2&Ensslin} and diffusion models \citep{CRE3&Ptuskin, CRE4&Breit,CRE6&Shlachi,CRE5&Dogiel}, which has been shown to vary galaxy by galaxy \citep{taba_13,nasirzadeh2024radio}. As the magnetic field structure also changes locally, the scale of the B/CR equipartition is also expected to change accordingly (i.e., in the disk and at the center). {\apjrv The} nonthermal correlation with the IR bands observed (Section~\ref{sec: radio-IR}) strengthens the possibility of the B/CR equipartition, in particular, in colder phases of the ISM away from winds/outflows.

By assuming the equipartition condition between the energy densities of the magnetic fields and cosmic rays ($\rm \epsilon_{CR}=\epsilon_{B} = B_{eq}^2/8\pi$), the total magnetic field strength can be calculated from the nonthermal (synchrotron) intensity, $\rm I_n$, by substituting $n_e$ in the cosmic ray energy density with the synchrotron intensity \citep{beck2005revised}.   
Based on \cite{beck2005revised}, the strength of the equipartition magnetic field in Gauss is given by:

\begin{equation}{\label{eq:mf}}
\rm B_{eq} = \left(\frac{4\pi(2\alpha_n+1) (K+1) I_n E_p^{1-2\alpha_n}\left(\frac{\nu}{2C_1}\right)^{\alpha_n}}{(2\alpha_n-1)C_2(\alpha_n)LC_4({\it i})} \right)^{\frac{1}{\alpha_n+3}} 
\end{equation}
where $\rm E_p$ represents the rest energy of the proton ($\rm E_p = 1.5\times10^{-3}$\,erg); {\apjrv $\alpha_n$ and $\nu$ are the nonthermal spectral index and frequency}; $i$ the inclination; and C1, C2, and C4 are combinations of physical constants (see \hyperref[sec:appendix b]{Appendix\,A}). {\apjrv In this equation, K represents the ratio between the number densities of cosmic ray protons and electrons, and L is the path length through the synchrotron emitting medium.}

 {\apjrv The} nonthermal map obtained in Section~\ref{sec.separation} is used for this purpose. In the absence of information on the magnetic field inclination, we adopt the inclination of the CND in Equation~(\ref{eq:mf}). The inclination of this region is reported to range from $65^{\circ}$ to $80^{\circ}$ \citep{martin2012surviving,hsieh2017molecular,goicoechea2018high}. We used $72^{\circ}$ as the median value in our calculations. The uncertainty in $\rm B_{eq}$ resulting from the inclination angle is roughly 20\%. In this study, we adopted the nonthermal spectral index of 0.7, as reported by \cite{reich1988study} and \cite{sato2024spiral}. In general, this parameter can change between  0.7 and 1 in the vicinity of the SMBH \citep{sato2024spiral}, which represents about 15\% uncertainty in $\rm B_{eq}$. In Equation~(\ref{eq:mf}), the frequency and K ratio are assumed to be 1.3\,GHz and 100. \cite{beck2005revised} mentioned that, for the K ratio, all current models of cosmic ray origin predict proton dominance. However, this ratio varies significantly in different regions and depending on the cosmic ray origin. For injection at 1\,TeV, \cite{crocker2011wild} finds acceptable fits to the data in the GC by assuming K=100. 
To determine the L parameter in Equation~(\ref{eq:mf}), the width of this central disk ranges from 8 to 14\,pc \citep{sutton1990co, oka2011new, blank2016inner}. The median value of 11\,pc\footnote{ This size is relevant to the overall dimensions of the circumnuclear region; however, it may not be suitable for discrete features like pulsar wind nebula \citep{wang2006g359} within this area, which will be addressed in a separate paper at a higher resolution in the future.} is adopted in our analysis. The resulting uncertainty in $\rm B_{eq}$ due to variations in L is estimated to be less than 10\%.

\begin{figure*}[!ht]
    \centering
    \includegraphics[width=11cm]{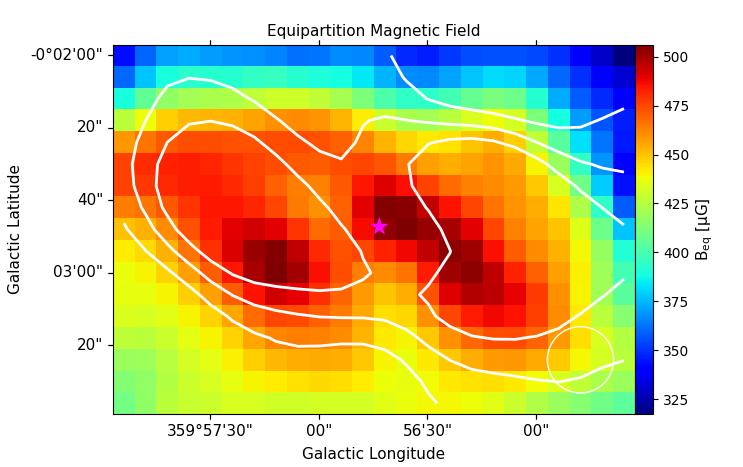}
    \caption{Equipartition magnetic field strength {\FT mapped} in the circumnuclear region (R$<7$\,pc) with contours of the 160\,$\mu$m emission tracing a relatively cool ISM. Colorbar represents the strength in units of $\mu$G. The beam size of $18\arcsec \times 18\arcsec$  is indicated in the lower right corner, and the star represents the position of Sgr\,A$^{\star}$. Contour levels are 40, 46, and 52\,Jy\,pixel$^{-1}$, respectively.}
    \label{mfmap}
\end{figure*}

{\apjrv Figure~\ref{mfmap}} shows that the equipartition magnetic field is strong {\apjrv at} the position of Sgr\,A$^{\star}$ ($\rm B_{eq}\simeq$ 502.6 $\pm$ 10.6\,$\mu$G at 18\arcsec~resolution). There are also two parallel patches of strong magnetic field ($\rm B_{eq}>450\,\mu$G) in the north and south parts of the CND, which overall have a good correspondence with the relatively cool gas distribution traced by the 160$\mu$m emission. 
We further compare the distribution of different ISM components around Sgr\,A$^{\star}$  in Section~\ref{sec:radial} in more detail. In the entire region of this study, the median value of the equipartition magnetic field is $\rm B_{eq}$ = 445.5$\pm$7.2\,$\mu$G.  The reported error {\apjrv for the median value is} derived by propagating the errors in the nonthermal intensity, taking into account the calibration and statistical uncertainties.

We studied the equipartition magnetic field within 7\,pc from the SMBH, while other {\apjrv works} calculated magnetic field strength in larger areas. Within a 50\,pc distance from the center, \cite{chuss2003magnetic} reported a magnetic field strength of less than 3\,mG using dust polarized emission at 350\,$\mu$m  and assuming a balance between kinetic and magnetic field energy densities. Also for a region of about 30\,pc around the center of the Galaxy, \cite{akshaya2024magnetic} reported that the mean strength of the B-field projected on the plane of the sky is $\sim$2\,mG. This employs the Davis-Chandrasekhar-Fermi (DCF) methods and utilizes thermal dust polarization observations to evaluate the strength of the magnetic field. Specifically in the CND, \cite{guerra2023strength} used the modified DCF method and 53\,$\mu$m polarimetric observations from SOFIA/HAWC+, estimating the plane-of-sky component of the B-field to be at least 1\,mG. Therefore, even after accounting for all uncertainties in our calculations, the obtained equipartition magnetic field in this study is generally lower than the reported magnetic field strength by other methods ($\rm B_{eq}<1\,mG$). Recently, \cite{yusef2022statistical} calculated the equipartition magnetic field strength in radio filaments of the CMZ to be on the order of $\sim$100–400$\mu$G. The combination of a low B and high cosmic ray pressure \citep{oka2005hot} suggests that equilibrium can be maintained in the central region. Moving forward, our main goal is to explore whether even this relatively weak magnetic field could play a crucial role in structure formation within the inner 7\,pc of the SMBH. Compared to other studies, \cite{larosa2005evidence} estimated a relatively low value of the magnetic field ($\simeq$10\,$\mu$G) using low-frequency RC data at 74 and 330\,MHz for diffuse nonthermal sources across the entire GC at a resolution of 125$^{''}$, which is likely due to absorption effects of the synchrotron emission at low frequencies. This is confirmed by the observed break in the radio spectrum by  \cite{crocker2010lower} who reported a lower limit of about 100\,$\mu$G over $\simeq400$\,pc scales.


\subsection{Thermal Ionized Gas Density}\label{sec:ionized}
\begin{figure*}[!ht]
    \centering
    \includegraphics[width=\textwidth]{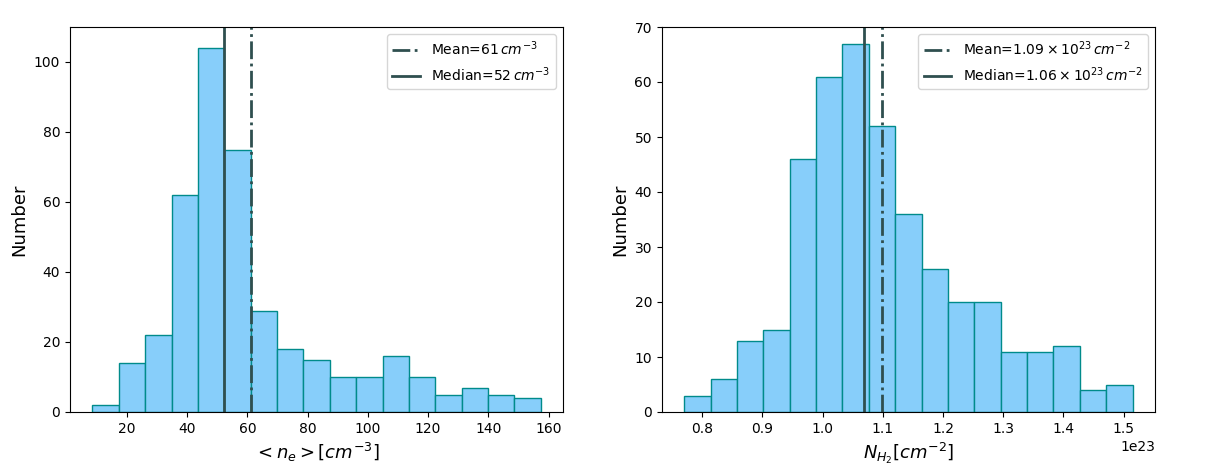}
    \caption{{\FT Histogram distributions} of $\rm N_{H_2}[cm^{-2}]$ and $\langle n_e\rangle[cm^{-3}]$, {\apjrv at the resolution of $18\arcsec$}, in the circumnuclear region within 7\,pc from the SMBH.}
    \label{hist}
\end{figure*}

The sub-millimeter RRLs can be a reliable tracer of thermal electrons in ionized regions.
To determine the thermal electron density in warm ionized gas, we used Equation~(\ref{eq:1}), which is applied to determine the EM. Given that $\rm EM = \int n_e^2 dl = \langle n_e^2\rangle$L, the volume-averaged thermal electron density ($\rm \langle n_e\rangle$) along the line of sight under optically thin conditions is: 

\begin{equation}\label{eq.ne}
 \rm \langle n_e\rangle[cm^{-3}] = \left[\frac{{\it f} \times (\frac{T_{L}\Delta v}{K\,kms^{-1}}) \times (\frac{T_e}{K})^{1.5}(\frac{\nu}{GHz})}{5.76\times 10^{2}(\frac{L}{pc})}\right]^{0.5}   
\end{equation}
where \( f \) denotes the volume filling factor, which quantifies fluctuations in \( n_e \), and is related to the mean electron density by \( \langle n_e \rangle = \sqrt{f \langle n_e^2 \rangle} \); its average value is approximately 0.2 \citep{berkhuijsen2006filling} in the warm ionized gas toward the center of the galaxy. As mentioned in Section~\ref{sec.separation}, $\rm T_{L}\Delta v$ represents the integrated line brightness temperature over the velocity width $\Delta v$, while $\rm T_e$ denotes the electron temperature. The H40$\alpha$ recombination line serves as the thermal tracer and the frequency ($\nu$) is 99.0\,GHz. In this equation, the path length (L) of the ionized gas, similar to Section~\ref{sec.mf}, is 11\,pc and $\rm T_e$ = 6000\,K. The mean volume-averaged thermal electron density in the inner 7\,pc of the GC region is 61.3\,$\pm$\,1.4\,cm$^{-3}$. The reported error for the mean value is calculated as the standard deviation divided by the square root of the sample size. The histogram on the left in Figure~\ref{hist} displays the distribution of thermal electron density.

Although the propagated error from the integrated line intensity is less than 10\%, the main source of uncertainty in Equation~(\ref{eq.ne}) arises from variations in electron temperature, diameter, and volume filling factor. The thermal electron density would not be changed by more than 15\% due to variations in both $\rm T_e$ and L. The reported volume filling factor is between 0.15 and 0.25 toward the center \citep{anantharamaiah1986ionized,berkhuijsen2006filling}, which leads to an uncertainty of less than 13\% in $\langle n_e\rangle$.

\subsection{Molecular Gas Column Density}\label{sec:density}

 The molecular gas column density $\rm N_{H_2}$ is obtained using the following equation:
 \begin{equation}
     \rm N_{H_2}[{\rm cm^{-2}}] = X_{\rm CO}[{\rm cm^{-2}(K\,km s^{-1})^{-1}}]\times I_{\rm CO}
 \end{equation}
 in which $\rm I_{\rm CO}$ is the integrated intensity of the CO line and $\rm X_{CO}$ is the $\rm CO$-to-$\rm H_2$ conversion factor. We use the mean value of $\rm X_{co}=0.48\pm0.15 \times 10^{20} cm^{-2}(\rm K\,km\,s^{-1})^{-1}$ in the GC, as reported by \cite{kohno2024co}. {\apjrv Due} to a higher excitation state, $^{12}$CO (J=3$ \rightarrow $2) traces a warmer and denser environment compared to $^{12}$CO (J=1$ \rightarrow $0) \citep{akshaya2024magnetic}. Additionally, because of better sampling of the CHIMPS2 survey, we used the J=3$ \rightarrow $2 transition to measure the molecular gas column density. It is important to note that the $\rm X_{CO}$ conversion factor is defined for $^{12}$CO (J=1$ \rightarrow $0), so the intensity of $^{12}$CO (J=3$ \rightarrow $2) from the CHIMPS2 survey needs to be converted to the J=1$ \rightarrow $0 rotational transition before the calculation. {\apjrv By} taking the ratio of $^{12}$CO~(J=3$\rightarrow$2) to $^{12}$CO~(J=1$\rightarrow$0) intensities in our studied region, we obtained a mean value of $0.70 \pm 0.01$, which is consistent with the conversion factor reported by \cite{tokuyama2019high} for this transition in the GC, $\rm \langle R_{3-2/1-0} \rangle = 0.70 \pm 0.06$. 
 We achieve an average {\apjrv molecular gas} column density of 1.09$\pm$0.01$\times 10^{23}\,\rm cm^{-2}$ in the circumnuclear region. {\apjrv The} error in the mean value of $\rm N_{H_2}$ represents the standard deviation divided by the square root of the sample size. The uncertainty in $\rm X_{CO}$ affects the mean value of $\rm N_{H_2}$ by 30$\%$, whereas the variation in $\rm N_{H_2}$ is less than 10$\%$ due to the uncertainty in $\rm \langle R_{3-2/1-0}\rangle$. The histogram on the right in Figure~\ref{hist} displays the distribution of the molecular gas column density in the CND region. {\apjrv In} this central region, the column density of molecular hydrogen gas has been reported to lie between $\mathrm{N}(\mathrm{H}_2) \sim 10^{22} - 10^{24}\,\mathrm{cm}^{-2}$ \citep{christopher2005hcn,requena2012great}, consistent with the values we obtained.


\section{Discussion}\label{sec.dis}

 In this section, we study the distribution of the obtained ISM components, compare the thermal and nonthermal energy densities, and investigate the role of the magnetic field and turbulence in controlling the ISM structures and star formation in the circumnuclear region. In addition, we study how the magnetic field affects molecular gas and its gravitational collapse.


\subsection{Distribution of the ISM components in vicinity of Sgr\,A$^{\star}$} \label{sec:radial}

\begin{figure*}[!ht]
	\centering
	\includegraphics[width=\textwidth]{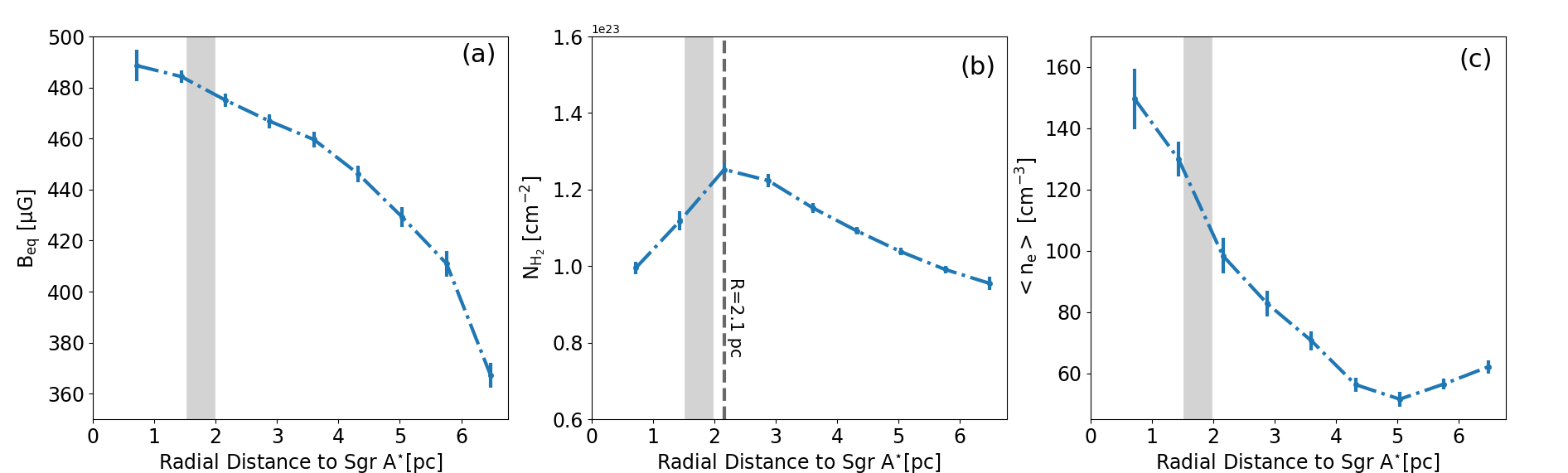}
        \includegraphics[width=\textwidth]{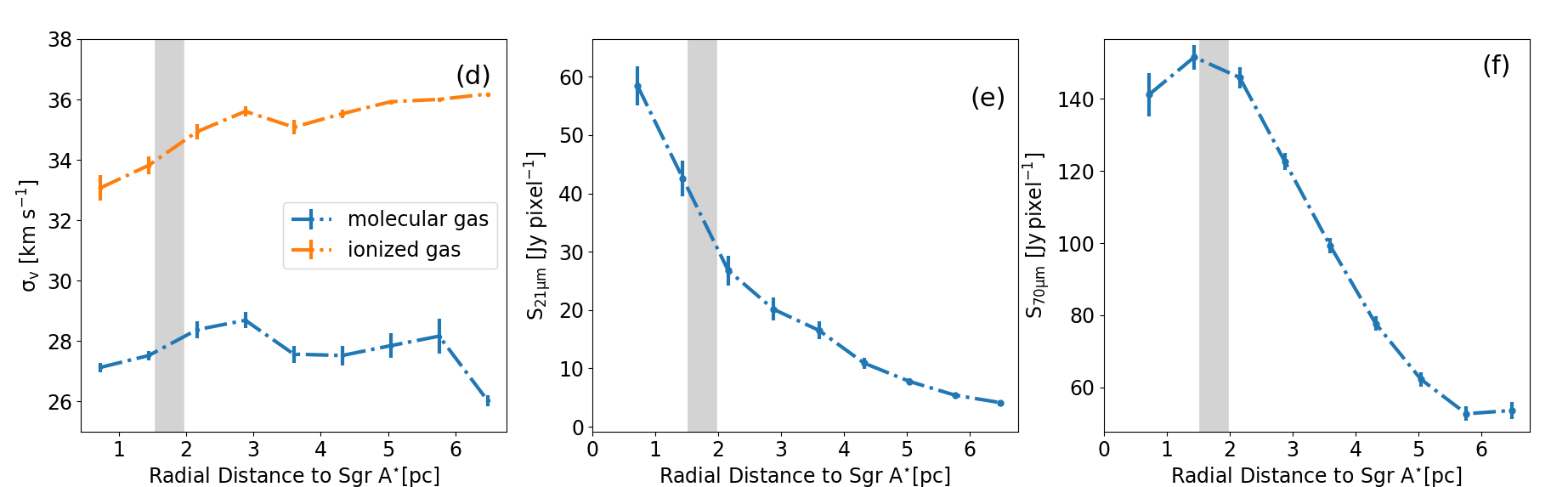}
	\caption{ Radial distribution of the physical properties of the ISM in the circumnuclear region (R$<$7\,pc):  ({\it a}) equipartition magnetic field strength ($\rm B_{eq}$), ({\it b}) $\rm H_2$ column density ($\rm N_{H_2}$),  ({\it c}) thermal electron density ($\langle n_e \rangle$), and ({\it d}) velocity dispersion of the ionized and molecular gas. Also shown are radial profiles of the ({\it e}) 21\,$\mu$m ($\rm S_{21\,\mu m}$) and ({\it f}) 70\,$\mu$m flux densities ($\rm S_{70\,\mu m}$). Points show mean values in rings of $18\arcsec$ width, and error bars show the standard deviation divided by the square root of the sample size in a ring. The gray shaded bars indicate the location of the inner radius of the CND based on different studies in the literature.}
	\label{fig:radial1}
\end{figure*}

The results obtained in Section\,\ref{sec:analysis} allow us to compare the distribution of the ionized and neutral phases of the ISM with those of the nonthermal ISM components, B/CREs, and turbulent gas velocity in the inner 7\,pc of the Milky Way. Figure~\ref{fig:radial1} shows  the 0.7\,pc (18$\arcsec$) ring mean radial profiles of equipartition magnetic field ($\rm B_{eq}$), molecular gas column density ($\rm N_{H_2}$),  volume-averaged thermal electron density ($\langle n_e \rangle$), and molecular {\apjrv and} ionized gas velocity dispersions  ($\sigma_v$). 

The inner edge of the CND {\apjrv is} at $\sim$1.5-2\,pc as reported by previous studies and using various line emissions \citep{jackson1993neutral,christopher2005hcn}. The gray shadow regions in radial profiles indicate this approximate boundary. The magnetic field increases toward the center while the molecular gas column density peaks at R$\sim$2\,pc and then decreases toward the center. Moving forward, we divide our analysis based on the molecular gas's peak because neutral molecular gas is a characteristic property of the ISM. Figure~\ref{fig:radial1} also indicates that the expected positive relationship between the magnetic field and molecular column density does not hold at a galactocentric radius smaller than {\apjrv $\sim$2\,pc} from the black hole.

The void of molecular gas column density within the inner radius of the CND, called the "ionized cavity" \citep{bryant2021episodic}, could be sustained by outflows generated by massive stars in the nuclear cluster \citep[e.g.,][]{genzel1994nucleus}. {\apjrv More reasons based on feedback from the SMBH} have been suggested by other studies. \cite{blank2016inner} used simulations to indicate that suppressive forces like magnetic fields {\apjrv near the SMBH} play a more significant role in maintaining this cavity than outflows from massive stars of the nuclear cluster in this region.
Moreover, this ionization of the ISM has been suggested by the association between the molecular filaments and the ionized mini-spiral as the gas approaches Sgr\,A$^{\star}$ \citep{martin2012surviving}. This cavity can also be caused by the SMBH's older AGN phase \citep{feruglio2010quasar} by removing the cold gas and/or ionizing it. Most galaxies have experienced the AGN phase, and AGN activities can play important roles in the formation and evolution of their host galaxies \citep{mo2010galaxy}. The Milky Way could have experienced this evolutionary phase a few million years ago. The presence of the Fermi bubbles both above and below the GC provides evidence of past AGN activity in the galaxy \citep{yang2018unveiling}.

As illustrated in Figure~\ref{fig:radial1}, the density of thermal ionized gas increases within 5\,pc of the SMBH. {\apjrv Further}, the radial distribution of velocity dispersion of $^{12}$CO (J=1$ \rightarrow 0)$, which is discussed in the next subsection in more detail, shows a roughly constant trend with two peaks around 3 and 6\,pc from the SMBH. {\apjrv The nearly uniform velocity dispersion indicates that the turbulent velocity remains almost constant within R$<$7\,pc of Sgr\,A$^{\star}$}. In addition to the molecular gas, we indicate the velocity dispersion of ionized gas from H40$\alpha$ RRL in this region.  Compared to the molecular gas, the ionized gas shows a greater and progressively decreasing velocity dispersion toward the center.

{\apjrv In} addition to the magnetic field, molecular gas, thermal ionized gas, and velocity dispersion, we examine the ring mean radial distribution of the 21\,$\mu$m and 70\,$\mu$m emission {\apjrv to investigate the distribution of warm and cold dust in this region in comparison to molecular gas.} 
Figure~\ref{fig:radial1} shows that the 70\,$\mu$m emission drops within 2\,pc of the SMBH, whereas the 21$\mu$m emission does not exhibit such a deficiency in the innermost region of this central area.

\begin{figure*}[!ht]
	\centering
        \includegraphics[width=0.9\textwidth]{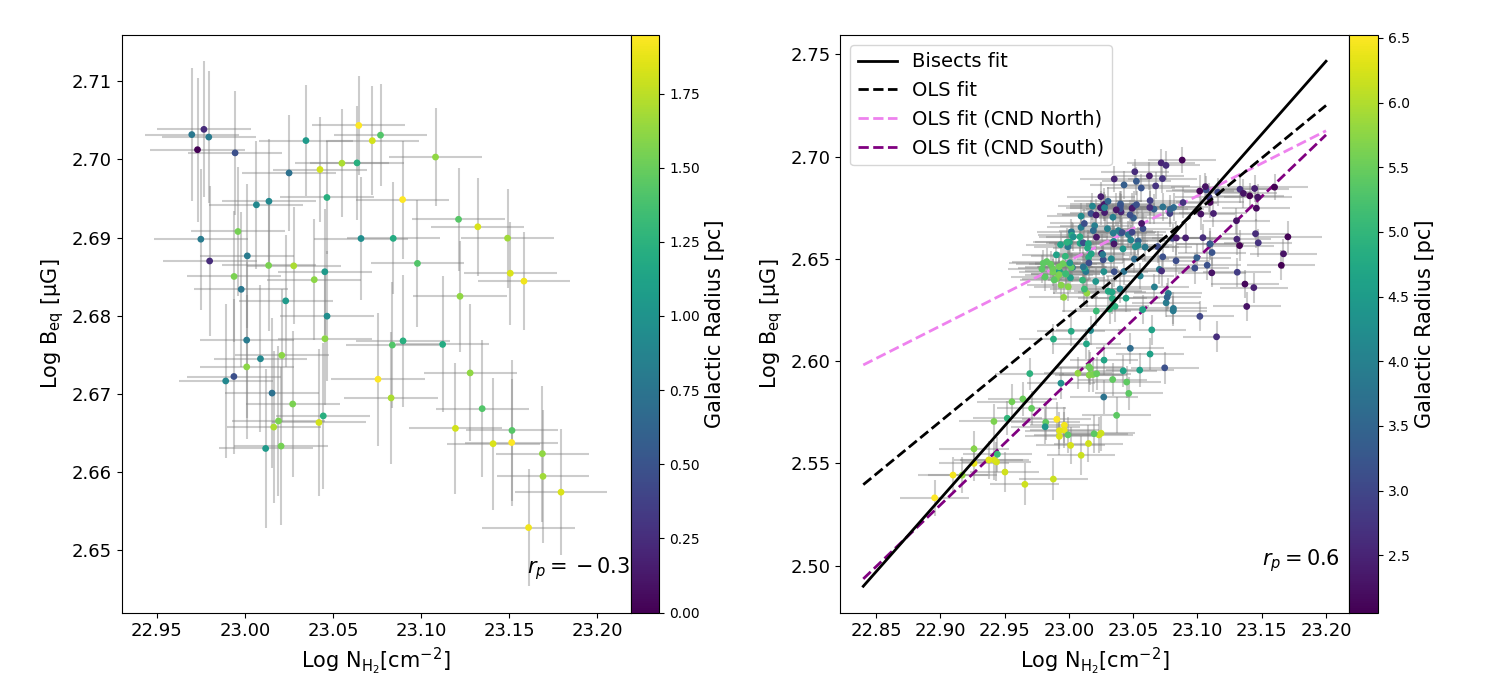}
        \includegraphics[width=0.9\textwidth]{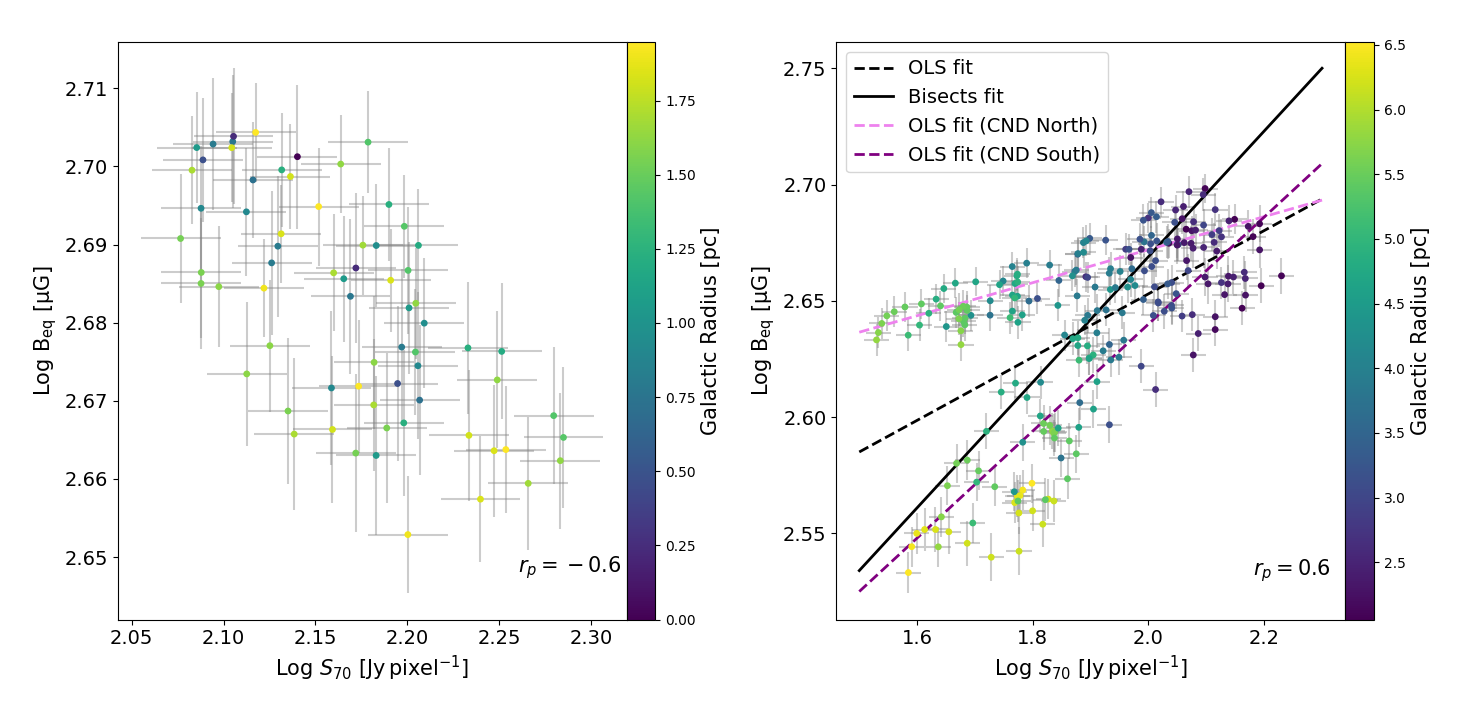}
	\includegraphics[width=0.9\textwidth]{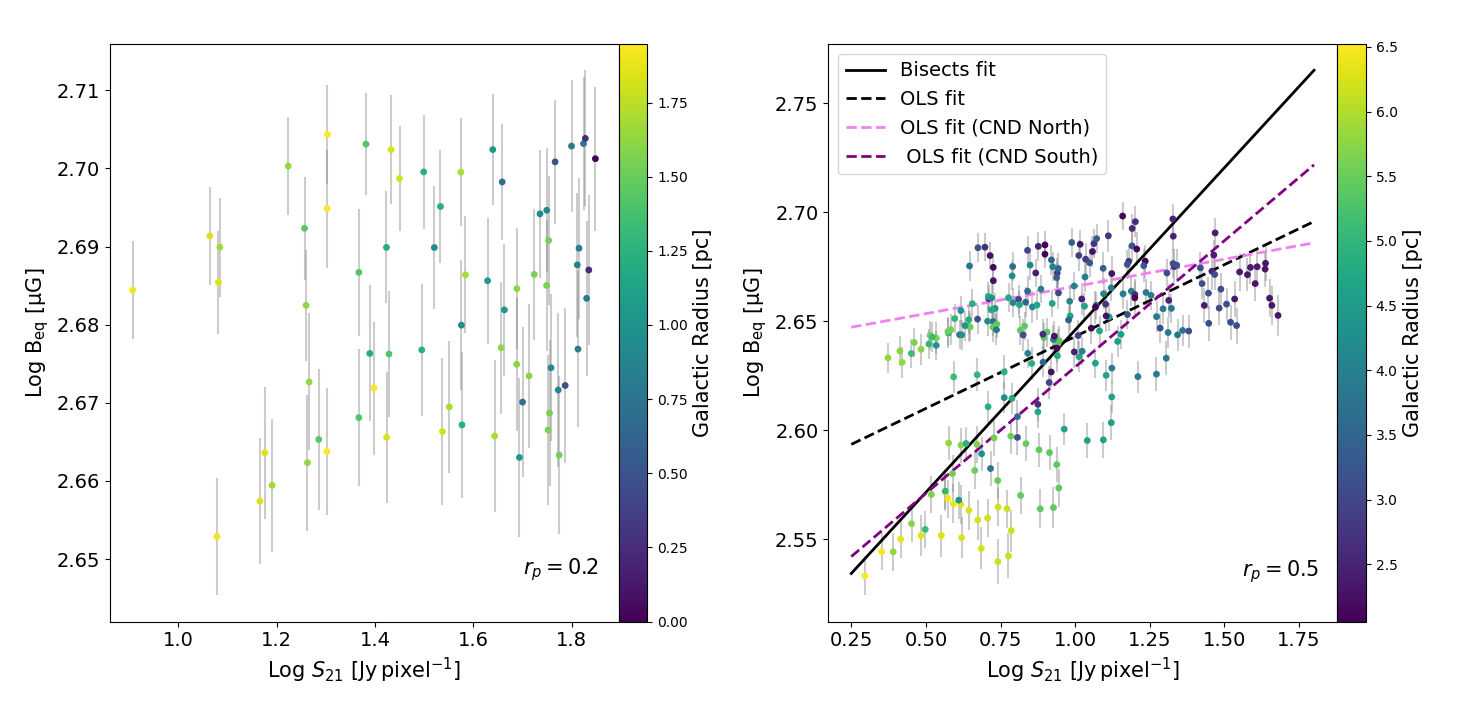}
        
	\caption{Relationships between the logarithm of equipartition magnetic field and $\rm N_{H_2}$ (top), 70\,$\mu$m (middle), and 21\,$\mu$m (bottom) flux densities in R$\le$2\,pc (left) and R$>$2\,pc (right) regions. Dashed black lines are OLS fits, while solid lines represent bisector fits. Dashed pink and purple lines show OLS fitted curves to the CND-north and CND-south, respectively (see Table~\ref{tab3}).}
	\label{fig:cor}
\end{figure*}

We further investigate the correlation between the equipartition magnetic field with molecular gas column density and the 21\,$\mu$m and 70\,$\mu$m emission in the circumnuclear region (Figure~\ref{fig:cor}). The left panels in Figure~\ref{fig:cor} show
the region within the radial distance of 2\,pc of Sgr\,A$^{\star}$ (R$\leq$2\,pc), and the right panels correspond to R$>$2\,pc in the CND. No significantly positive correlations are detected between $\rm B_{eq}$ and molecular gas and warm dust in the R$\leq$2\,pc region. The correlations are even negative (anti-correlation) with $\rm N_{\rm H_2}$ and 70\,$\mu$m emission, which is a result of their inverse radial trends in this region (Figure~\ref{fig:radial1}); while the molecular gas and the 70\,$\mu$m emission decrease toward Sgr\,A$^{\star}$, the equipartition magnetic field increases. This is likely due to feedback from the SMBH that ionizes gas, heats dust, and amplifies magnetic fields (and/or increases the number density of CREs). On the other hand, the absence of a positive correlation is not in favor of an equipartition between the magnetic field and CREs at R$\leq$2\,pc and hence $\rm B_{eq}$ does not represent the true magnetic field strength there.
 This observation unveils the abnormal characteristic of the magnetized ISM in the vicinity of the Milky Way's SMBH.

As shown in Figure~\ref{fig:cor}-right panels, positive correlations hold, particularly, between $\rm B_{\rm eq}$ and $\rm N_{\rm H_2}$ {\apjrv indicating} that, at the resolution of 0.7\,pc, the equipartition magnetic field is associated with cold molecular gas in the main area of the CND at R$>$2\,pc as suggested by the radio--FIR correlation (Section~\ref{sec: radio-IR}). A detailed discussion of very dense and cold molecular structures at higher resolutions, however, lies beyond the scope and available data of this paper. {\apjrv As shown in Table~\ref{tab3},} the slopes obtained from the OLS and bisector fits of $\rm B_{eq}$ versus $\rm N_{H_2}$ are 0.52$\pm$0.05 and 0.71$\pm$0.05, respectively, suggesting that the magnetic field may play a significant role in regulating star formation within the CND \citep{crutcher1996hi}. 

\begin{table}
	\centering
	\caption{ Logarithmic correlations and best fits of $\rm B_{eq}$ with $\rm N_{H_2}$, $\rm S_{70\,\mu m}$, and $\rm S_{21\,\mu m}$ for R $>$2\,pc.}
	\begin{tabular}{cccc}
		\hline\hline
             Correlations&$m_{\rm OLS}$\footnote{ Power-law slope of the relations using OLS fitting} &$m_{\rm bis}$\footnote{ Power-law slope of the relations suing bisector regression}&
            $r_p$\footnote{Pearson correlation coefficient}\\
		\hline
             $\rm B_{eq}-N_{H_2}$&0.514$\pm$0.050&0.713$\pm$0.045&0.56$\pm$0.04\\
             $\rm B_{eq}-S_{70\mu m}$&0.135$\pm$0.014&0.266$\pm$0.013&0.56$\pm$0.05\\
             $\rm B_{eq}-S_{21\mu m}$&0.066$\pm$0.007&0.150$\pm$0.009&0.52$\pm$0.05\\
             \hline
             &\multicolumn{2}{c}{CND South}&\\
             $\rm B_{eq}-N_{H_2}$&0.603$\pm$0.052&0.739$\pm$0.054&0.74$\pm$0.05\\
             $\rm B_{eq}-S_{70\mu m}$&0.230$\pm$0.011&0.256$\pm$0.013&0.88$\pm$0.02\\
             $\rm B_{eq}-S_{21\mu m}$&0.116$\pm$0.009&0.151$\pm$0.011&0.83$\pm$0.03\\
             \hline
             &\multicolumn{2}{c}{CND North}&\\
             $\rm B_{eq}-N_{H_2}$&0.318$\pm$0.046&0.358$\pm$0.046&0.66$\pm$0.07\\
             $\rm B_{eq}-S_{70\mu m}$&0.071$\pm$0.004&0.071$\pm$0.004&0.89$\pm$0.02\\
             $\rm B_{eq}-S_{21\mu m}$&0.025$\pm$0.005&0.047$\pm$0.005&0.54$\pm$0.08\\
		\hline
           
	\end{tabular}
	\label{tab3}
\end{table}

Studies in nearby galaxies show that the equipartition magnetic field strength from synchrotron emission data is correlated with the SFR surface density $\Sigma_{\rm SFR}$ following a power-law relation $B\propto \Sigma_{\rm SFR}^b$ with $b \leq 0.3$ \citep[][]{Chyzy,tabatabaei2013detailed,Heesen,hassani2022role,tabatabaei2022cloud}. {\apjrv Theoretically}, an exponent of $b=0.34\pm 0.08$ is expected 
if a supernova induced small-scale turbulent dynamo in star-forming regions amplifies the magnetic field \citep{gressel,schleicher2013new}. In general, \cite{barnes2017star} used 24 and 70\,$\mu$m emissions to estimate the SFR in the CMZ by SFR calibration relations \citep{kennicutt2012star,calzetti2010calibration}. {\apjrv These} calibration relations rely on a large, complete stellar population spanning various evolutionary stages and are not applicable to smaller structures \citep{kennicutt1998star,murphy2011calibrating}. For this reason, for the linear scales relevant for this study, we can not use these relations to estimate the SFR surface density. Therefore, we focus on the relationship between $\rm B_{eq}$ and the 70 and 21\,$\mu$m emissions to determine which of these star formation tracers exhibit a slope close to 0.3 in the CND. {\apjrv As presented in Table~\ref{tab3},} the equipartition magnetic field shows a flat relationship with the 21\,$\mu$m emission with slopes of 0.07 (OLS) and 0.15 (bisects). In contrast, the slopes for the 70\,$\mu$m emission, at 0.14 (OLS) and 0.27 (bisects), are closer to the theoretically expected value. As discussed in Section~\ref{sec: radio-IR}, the 21\,$\mu$m emission can be linked to sources different from dust in the ISM and young massive stars. In particular, 
in the vicinity of the SMBH, besides the presence of young massive stars \citep{yusef2013alma,yusef2015signatures}, there are strong ionizing shocks that can heat the dust re-radiating in mid-IR. Thus, it is unlikely that the 21.3\,$\mu$m emission in the CND is solely caused by current star formation activity. {\apjrv Similar to the radio--IR correlations discussed in Section~\ref{sec: radio-IR}, the correlations at R$>$2\,pc follow different trends in the north and south of the CND (shown by pink and purple dashed lines in Figure~\ref{fig:cor}). As discussed in Section~\ref{sec: radio-IR}, the flatter slopes of the relationships in the north (Table~\ref{tab3}) can be linked to the winds and outflows in this region \citep{zhao2016new}.}

\subsection{ISM Energy Balance at the Center}\label{sec.balance}

\begin{figure}
	\centering     
	\includegraphics[width=\columnwidth]{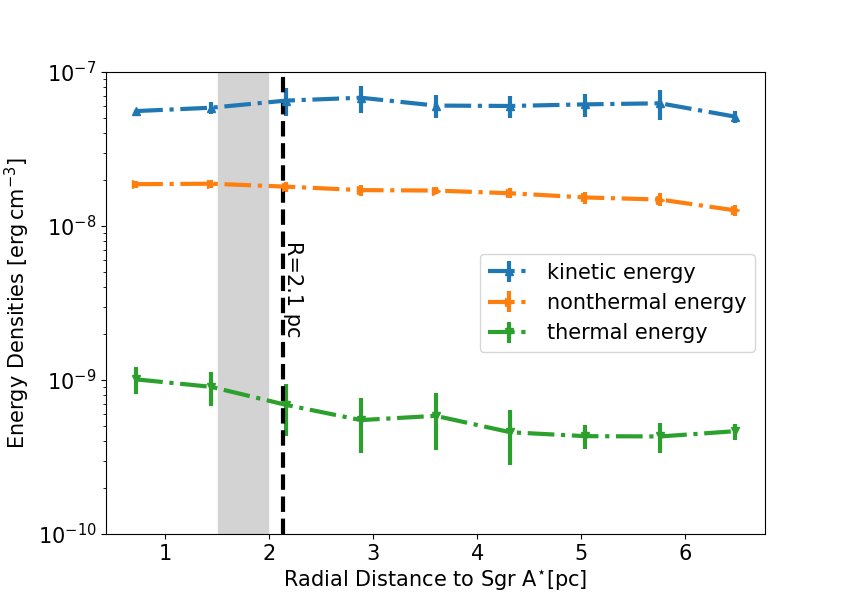}
	\caption{Radial distribution of the ISM energy densities and their variations with galactocentric radius in the circumnuclear region. The gray shaded bar indicates the location of the inner radius of the CND based on different studies in the literature. The dashed line represents the obtained boundary by $^{12}$CO (J=3$\rightarrow$2) in Figure~\ref{fig:radial1}.
     }
 \label{fig:eb}
\end{figure}

 Investigating the energy balance of the ISM in the inner 7\,pc of the SMBH is essential for a deep understanding of its physical conditions. The kinetic energy density of turbulent gas motions, the thermal energy of different gas phases, and the nonthermal energy densities from the magnetic field and cosmic rays \citep{tabatabaei2018discovery} shape the total energy content of the ISM. The obtained physical characteristics of the ISM allow us to compare the thermal and nonthermal energy densities in the circumnuclear region (see Figure~\ref{fig:eb}). Estimating each energy density is described {\apjrv in the following.}

The kinetic energy density of turbulence is given by $\rm E_{k}=\frac{1}{2}\rho \sigma_{v}^2$, where $\rho$  is calculated for the 11\,pc scale structures from the molecular mass surface density ($\rho=\frac{\Sigma_{\rm H_2}}{11\,\rm pc}[\rm M_{\odot}\,pc^{-3}]$) obtained using the CO-to-$H_2$ conversion factor ($\alpha_{co} = 1.1\,\rm[M_{\odot}\,pc^{-2} (K\,kms^{-1})^{-1}]$) and the intensity of CO emission ($\Sigma_{\rm H_2}=\alpha_{co} \rm I_{co} [M_{\odot}\,pc^{-2}]$ ). The value of $\alpha_{co}$ for the center of the Milky Way is obtained by dividing $\rm X_{co}$ {\FT (see Section~\ref{sec:density})} by a factor of 4.5 $\times 10^{19}$ $[\rm cm^{-2}(M_{\odot}\,pc^{-2})^{-1}]$ \citep{teng2022molecular}. The turbulent velocity ($\sigma_v$) is determined by $\sigma_v=\delta V_{int}/\sqrt{8\ln{2}}$, where $\delta V_{int}$ represents the velocity dispersion of  the CO line width. To determine the kinetic energy, we employed the zero and second moment maps of  $^{12}$CO (J=1$\rightarrow$0) of the NRO 45m telescope \citep{tokuyama2019high} for calculating $\Sigma_{\rm H_2}$ and $\delta V_{int}$.  {\FT This energy density 
is also obtained for the ionized gas 
using the zero- and second-moment maps of the H40$\alpha$ RRL data to estimate its density (see Section~\ref{sec:ionized}) and velocity dispersion, respectively}. {\FT Compared to the molecular gas, the ionized gas has a much lower gas density and a comparable velocity dispersion (see Figure~\ref{fig:radial1}). As a result, the ionized gas contributes insignificantly to the kinetic energy density of the ISM.} 
For this reason, the kinetic energy density of the ionized gas is not considered. {\FT We also note that the contribution of the thermal line broadening is negligible in the velocity dispersions considered ($\sigma_{v}^{2}= \sigma_{v,\mathrm{nt}}^2 + (\rm \frac{kT}{m})^2$) for both molecular gas \citep[$\rm T=$\,63\,K,][]{oka2011new}  and ionized gas \citep[$\rm T=$\,6000\,K,][]{scoville2003hubble} and $\sigma_v$ is almost equivalent to that of the nonthermal broadening ($\sigma_v \simeq \sigma_{v,{\rm nt}})$.} 
{\FT Due to the absence of high-resolution 
HI data in the region of study (at lower than 1\,pc resolution)}, we could not calculate the kinetic energy density for the neutral atomic gas. However, the gas in the GC exhibits much higher surface density and is predominantly molecular \citep{morris1996galactic,kruijssen2014controls}. Additionally, according to \cite{sofue2022three}, the atomic gas mass in the GC is only 10\% of the GC's molecular mass. This suggests that neglecting the contribution of atomic gas could lead to an uncertainty in our results of at least 10\%. However, we don't have access to the velocity dispersion of atomic gas in the central region at high resolution. We should add that this uncertainty would be higher within the ionized cavity.

The thermal energy density of the ionized gas can be calculated for both hot ($\rm T\sim 10^6$\,K) and warm ($\rm T\sim 10^4$\,K) gas. To determine the thermal energy density for the warm ionized gas, we use $\rm E_{th}=2.1\times nKT$ formula \citep{ferriere2001interstellar} and the volume-averaged thermal electron density of the warm ionized gas ($\langle n_e \rangle$), assuming $\rm T_e = 6000$\,K  \citep{scoville2003hubble} as it is also assumed in Section~\ref{sec:ionized}. Two simplifying assumptions for calculating the thermal ionized gas are considered. First, we assume equilibrium between the energy densities of electrons and ions due to thermal equilibrium \citep{draine2010physics}, and consider the same thermal energy density for ions and electrons in our calculations. Second, the energy density of hot gas in the ISM of the Milky Way is comparable to or slightly greater than that of the warm gas, depending on its volume filling factor \citep{ferriere2001interstellar,beck2016magnetic}. In addition, the thermal energy density of both the molecular and atomic gas should be considered to evaluate the neutral gas using the $\rm E_{th}=1.1\times n KT$ formula \citep{ferriere2001interstellar}. We determined the thermal energy density of the molecular gas by using the molecular gas density ($\rho$) derived from the $^{12}$CO (J=1$\rightarrow$0) map and the temperature of 63\,K for the molecular medium in this central region \citep{oka2011new}. For the atomic neutral gas, the thermal energy density of both warm ($\rm T\sim5000$\,K) and cold ($\rm T\sim100$\,K) phases \citep{draine2010physics} is considered to be in thermal energy equilibrium \citep{ferriere2001interstellar}. As mentioned, due to the lack of a high-resolution HI map of the central region to calculate the warm neutral gas density, we {\FT indirectly estimated the average column density of atomic gas using the dust extinction ($\rm A_v$-to-$\rm N_H$ ratio) as dust is well-mixed with the gas \citep{bohlin1978survey,rachford2008molecular}. On average, the thermal energy density of the atomic gas ($\simeq 10^{-11}\,\rm erg\ cm^{-3}$) is one order of magnitude smaller than that of the ionized gas derived in this section ($10^{-10}\,\rm erg\ cm^{-3}$) and, therefore, it is neglected in this study.}

The nonthermal energy density of the magnetic field and cosmic-ray electrons is given by: $\rm E_{nt}=2E_b=\frac{B^2}{4\pi}$. The factor of 2 is needed because of the equipartition assumption. We use the equipartition magnetic field strength from Section~\ref{sec.mf} to calculate the lower limit to nonthermal energy density. This assumption can be the source of uncertainty in the nonthermal energy density calculation.  However, the correlation between synchrotron and FIR emission shown in Figure~\ref{fig:FIR-nth} suggests that this assumption can be generally valid in our study. We should note that this calculation is less certain for R$\leq$2\,pc.

In this investigation, we {\FT computed the thermal, nonthermal (magnetic field/cosmic rays), and kinetic (turbulent)} energy densities in the circumnuclear region. {\FT As shown in Figure\,\ref{fig:eb}, the nonthermal and turbulent energy densities dominate significantly over the thermal energy density at all radial distances from Sgr\,A$^{\star}$ (R$<$7\,pc), indicating their important role in structure formation in this central region.}
Based on our calculations, the mean values of the thermal, nonthermal, and kinetic energy densities in the circumnuclear region are $5.7\pm0.7\times 10^{-10} \rm erg\, cm^{-3}$, $1.6\pm0.1 \times10^{-8} \rm erg\, cm^{-3}$, and $6.1\pm0.3 \times10^{-8} \rm erg\ cm^{-3}$ respectively. The thermal energy density of the central region could be attributed to ionizing effects such as star formation activity {\apjrv in the center or radiation from Sgr\,A$^{\star}$}.  
Generally, the ISM in this region is characterized by a supersonic ($\rm E_k/E_{th}\gg 1$) and low-$\beta$ plasma  ($\rm \beta= E_{th}/E_b\ll 1$).  Similar conditions were also reported at the center of NGC~1097 by \cite{tabatabaei2018discovery}.

\begin{table}[ht]
\caption{The mean energy densities as well as $\beta$, $\sigma_v$, $V_A$, and $M_A$, in R$\leq$ 2 and R$>$ 2\,pc.}
    \centering
    \begin{tabular}{ccccc}
    \hline \hline
         &Properties &Units& R$\leq$ 2\,pc&R$>$2\,pc\\
    \hline    
         &$E_{th}/10^{-10}$ &$\rm[erg\,cm^{-3}]$&8.5$\pm$0.4&5.1$\pm$0.1\\
        
         &$E_{nt}/10^{-8}$ &$\rm[erg\,cm^{-3}]$&1.9$\pm$0.1&1.5$\pm$0.1\\
          
         &$E_{k}/ 10^{-8}$ &$\rm[erg\,cm^{-3}]$&6.1$\pm$0.2&6.1$\pm$0.1\\
        
         &$\beta/10^{-2}$ &-&9.2$\pm$0.5&6.6$\pm$0.1\\
          
         &$\sigma_v$ &$\rm [Km\, s^{-1}]$&27.9$\pm$0.2&27.6$\pm$0.1\\
         
         &$V_A$ &$\rm [Km\, s^{-1}]$& 10.9$\pm$0.1&9.9$\pm$0.1\\
         
         &$M_A$ &-&4.4$\pm$0.1&4.8$\pm$0.1\\
      \hline   
    \end{tabular}
    \label{table:2}
\end{table}

In addition to energy densities, we have assessed the Alfv\'{e}n velocity ($V_A = B/ \sqrt{4\pi\rho}$) and Alfv\'{e}n Mach number ($M_A =\sqrt{3}\sigma_v/V_A$) to investigate the motions within {\apjrv this region.} Our calculated Alfv\'{e}n velocity is close to what is reported by \cite{lu2024magnetic} for some specific molecular clouds in {\apjrv Sgr\,A (except for the CND), Sgr\,C, and dust ridge regions}. In the circumnuclear region, we observed Alfv\'{e}nic supersonic motions, where the Alfv\'{e}n velocity ($V_A$) exceeds the sound speed, which can be calculated as $\rm c_s = (kT_K/2.33m_H)^{1/2}$, where $\rm T_K$ represents the kinetic temperature and $m_H$ represents the mass of a hydrogen atom. By using the obtained kinetic temperatures in the GC molecular clouds and the CND, which fall between 25 to 200\,K \citep{huttemeister1993kinetic,oka2011new}, it is found that the Alfv\'{e}n velocity is at least one order of magnitude greater than the sound speed. Additionally, {\FT the high Mach number ($ M_A > 1$) obtained} in this central region aligns with previously reported high Mach numbers generally in the GC \citep{ morris1996galactic,ginsburg2016dense,akshaya2024magnetic}. {\apjrv It is worth mentioning that, since we adopted the equipartition magnetic field, the resulting Mach number is likely overestimated.} The mean values of calculated parameters in this subsection are reported in Table~\ref{table:2} for radial distances R$\leq$2\,pc and R$>$2\,pc from Sgr\,$\rm A^{\star}$. These values suggest that all properties, except for {\FT the thermal energy density — and consequently, 
$\beta$ — remain largely unchanged between the inner and outer 2\,pc} regions. The mean value of $\rm E_{th}$ in the inner ionized region is approximately twice as high as in the outer region.

In this subsection, our main conclusions remain unchanged despite variations and uncertainties in the conversion factors, length of the region, electron temperature, volume filling factor, and other assumed parameters. Additionally, by considering the uncertainties in the magnetic field strength described in Section~\ref{sec.mf}, the nonthermal energy density remains {\FT higher} than the thermal energy density. The two mentioned assumptions (equilibrium between the energy densities of ions and electrons, as well as hot and warm gas) in calculating the thermal energy density are typically applicable to the ISM of galaxies. {\FT More detailed studies of} the ISM of the circumnuclear region in the future will enable us to analyze the thermal density of this region with {\FT a better accuracy. For instance, the energy density of the soft X-ray plasma, which is the tracer of thermal hot ionized gas, is $3\times 10^{-10}\,\rm erg\ cm^{-3}$, as calculated by \cite{muno2004diffuse} , for the Sgr\,A region.} This value is close to the energy density of the thermal warm ionized gas we obtained for our studied region, which is $3.9\times 10^{-10}\,\rm erg\ cm^{-3}$. Additionally, \cite{muno2004diffuse} reported that in the vicinity of Sgr\,A$^{\star}$, the energy density of the hard X-ray plasma, which originates from nonthermal processes, is {\FT higher} than that of the soft X-ray plasma, which serves as a tracer for thermal hot gas and confirms the result of this subection ($\rm E_{th}<E_{nt}$). {\apjrv Another} source of uncertainty that should be mentioned is the impact of multi-velocity components along the line of sight, bulk motions (e.g., rotation, infall), and outflows on $^{12}$CO that can overestimate the obtained velocity dispersion derived from the second moment map. Nevertheless, our reported $\sigma_v$ is comparable to or slightly lower than {\FT that} 
obtained from C$^{18}$O observations with the elimination of minor components by \cite{yang2025jcmt}. In conclusion, the ISM in the vicinity of Sgr $\rm A^{\star}$ is affected by nonthermal and turbulent pressures, which {\FT can control} 
the star formation process, even under the equipartition assumption {\FT (also see Section~\ref{sec:mu})}.

\subsection{Mass-to-Magnetic Flux Ratio} \label{sec:mu}

In Sections~\ref{sec.mf} and \ref{sec:density}, we obtained the equipartition magnetic field strength ($\rm B_{eq}$) and the molecular gas column density ($\rm N(H_2)$). 
A classical approach to assess the impact of the magnetic field on gravitational collapse and star formation is investigating the ratio of the gas mass to the magnetic flux, $\rm M/\Phi_{B}$ \citep{strittmatter1966gravitational}.    
The magnetic flux is the magnetic field strength times the spatial area ($\rm \Phi_{B}=|B|\times$area) and the mass contained within this area can be estimated based on the measured molecular column density, assuming a 10\% helium content ($\rm M =2.8~N(H_2)~m_H\times$area) \citep{crutcher1999magnetic}. This parameter is normalized to the critical value ($\mu_0=(2\pi \sqrt{G})^{-1}$)  for an isothermal gaseous disk and derived by \cite{nakano1978gravitational} with linear perturbation analysis. The mass-to-magnetic flux ratio in units of the critical value ($\mu$) can be calculated with the formula provided by \cite{crutcher2004scuba}:

\begin{equation}
    \mu =7.6\times 10^{-21} \frac{\rm N(H_2)}{\rm B_{tot}},
\end{equation}

Here, $\rm B_{tot}$ represents the total magnetic field strength within the molecular cloud, measured in units of $\mu$G. {\FT The equipartition magnetic field (Section~\ref{sec.mf}) provides a lower limit for the actual magnetic field strength in this equation \footnote{because CREs can decouple the magnetic field due to winds and propagation effects.}}. Focusing on a purely molecular gas, $\rm N(H_2)$ denotes the molecular column density, measured in units of $\rm cm^{-2}$. The intrinsic value of $\mu$ ($\mu_{int}$) is determined by the ratio of $\rm N_{\perp}/B_{tot}$, where $\rm N_{\perp}$ is column density perpendicular to the sheet or disk geometry of matter in a flux tube, in the case of a magnetically supported cloud. For the median value, $\rm N_{\perp} =\frac 12 N(H_2)$ {\FT meaning} that $\mu_{int}$ =  $\frac12\mu$ \citep{heiles2005magnetic}.

The molecular clouds are characterized as critical if $\mu_{int}=1$ and subcritical when  $\mu_{int} < 1$, in which case the presence of a strong magnetic field prevents the gravitational collapse of the clouds. On the other hand, if $\mu_{int} > 1$, the clouds are known as supercritical and unstable to gravitational collapse, and hence the star formation can proceed to occur \citep{crutcher1999magnetic,heiles2005magnetic}. Considering more realistic situations, in which the magnetic flux is differently loaded with mass in a small cylindrical region, \cite{mouschovias1991magnetic} obtained a larger critical value of $\mu_{int}\simeq2$.

\begin{figure*}[ht!]
	\centering     
	\subfigure{\includegraphics[width= 11cm]{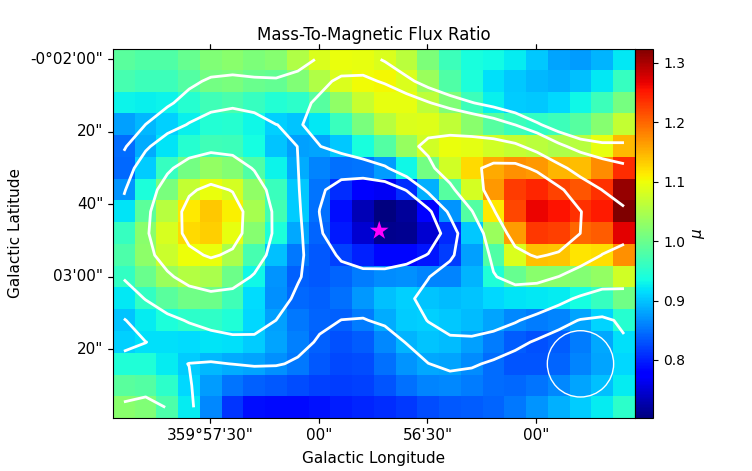}}
	\caption{Map of the mass-to-magnetic flux ratio ($\mu$) in the circumnuclear region with contours of the$^{12}$CO (J=3$\rightarrow$2) emission. The beam size of $18\arcsec \times 18\arcsec$ is shown in the lower right corner, and the star represents the position of Sgr\,A$^{\star}$. Contour levels are 1500, 1600, 1800, and 2000\,K km s$^{-1}$, respectively.}
        \label{fig:mu}
\end{figure*}

Figure~\ref{fig:mu} shows the distribution and variation of $\mu_{int}$ 
over the circumnuclear region. We find that $\mu_{int}<2$ with a median value of 0.92$\pm$0.02. Thus, even with a lower limit for the magnetic field strength, the ISM  obeys a subcritical condition. The maximum value of $\mu_{int}$ is 1.32$\pm$0.04 in the dense molecular part, and the minimum is 0.70$\pm$0.02 near Sgr\,A$^{\star}$. The reported errors are solely the propagated errors in the observed intensities. As discussed in Section~\ref{sec.mf}, the uncertainties in the assumed parameters for the magnetic field strength do not significantly impact the result of this subsection. Their impact is less than 20\% on this parameter. On the other hand, the median value of $\mu_{int}$ {\FT can change} by 30\% due to uncertainty in the $\rm X_{CO}$ conversion factor. 
To summarize, the mostly subcritical condition ($\mu_{int}\lesssim1-2$) remains intact taking into account the uncertainties in calculating the equipartition magnetic field and molecular gas column density. Using thermal dust polarization at 53\,$\mu$m, \cite{akshaya2024magnetic} found a lower mass-to-magnetic flux ratio than that obtained here because of their stronger magnetic field estimate.

\begin{figure}[ht!]
	\centering     
	\subfigure{\includegraphics[width=\columnwidth]{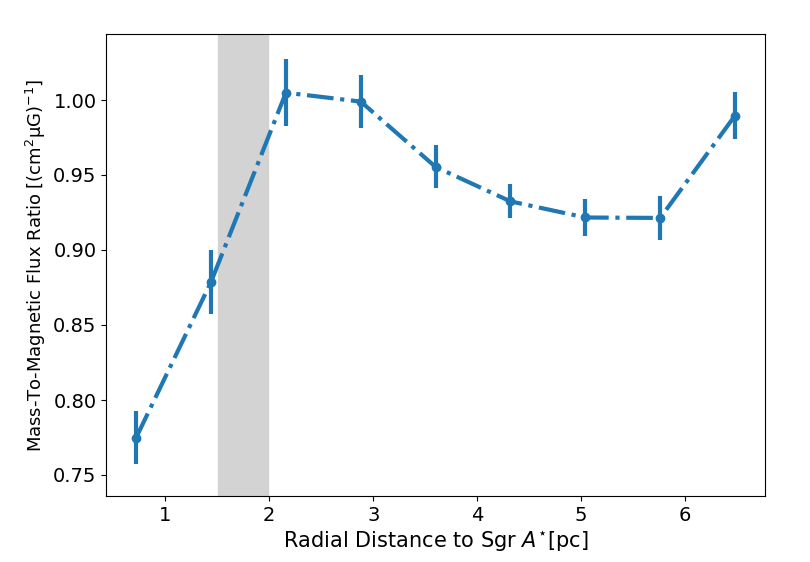}}
	\caption{\FT Radial distribution of the mass-to-magnetic flux ratio ($\mu$) in the circumnuclear region. Points show mean values in rings of
$18\arcsec$ and errors are the standard deviation divided by the square root of the sample size in a ring.  The gray shaded bar indicates the location of the inner radius of the CND based on different studies in the literature.}
        \label{fig:mu_radial}
\end{figure}

Figure\,\ref{fig:mu_radial} {\FT shows} the radial profiles of $\mu_{int}$. The mass-to-magnetic flux ratio increases from {\FT subcritical values of $\mu_{int}<0.8$ near} the center to {\FT $\mu_{int}>0.95$ at R=\,2.1\,pc, where the molecular gas density is highest (see Figure~\ref{fig:radial1}). At larger radial distances, $\mu_{int}$ changes insignificantly between 0.9 and 1.}
{\FT As discussed in Section~\ref{sec:radial}, the equipartition assumption underestimates effectively the magnetic field strength at R$\leq$2\,pc. Hence, the true mass-to-magnetic flux ratio, must be much smaller than those estimated here in this region.}

\section{Conclusions}\label{sec.conclusion}

To study the energy balance, structure formation, and physical condition in the ISM of the {\FT circumnuclear region (within 7\,pc distance from Sgr\,A$^{\star}$)}, we separate the thermal and nonthermal {\FT radio continuum (RC) components} at 1.3\,GHz using the H40$\alpha$ RRL as an extinction-free tracer of the free-free emission. {\FT The resulting RC maps are then used to 1) investigate correlations and mixing of different processes in the multi-phase ISM, 2) estimate the strength of the magnetic field and density of thermal electrons, and 3) obtain energy densities and map the mass-to-magnetic flux ratio in this region.} 
Our main findings are summarized as follows.

\begin{itemize}
    
    \item The nonthermal component dominates the RC emission at 1.3\,GHz everywhere in the GC. The mean value of the thermal fraction is $f^{\rm 1.3\,GHz}_{th}=13.1 \pm 0.2$\% at $4\arcsec$ resolution. The thermal emission is mainly emerging from the mini-spiral structure while the nonthermal component extends into a larger structure. 

    \item The nonthermal synchrotron emission is generally correlated with the IR emission. The correlation is almost linear with the 160\,$\mu$m emission. The weakest synchrotron--IR correlation holds with the 21\,$\mu$m emission in the north where strong winds of ionized gas exist. These indicate a fine balance between magnetic fields, CREs, and gas pressures in the colder phase of the ISM (particularly in south) which is confirmed by the correlation between the equipartition magnetic field B$_{\rm eq}$ and N$_{\rm H_2}$ in the R$>$2\,pc region (Figure~\ref{fig:cor}).      
    
    \item In the R\,$\le$\,2\,pc region, B$_{\rm eq}$ likely  deviates from the actual field strength as no positive B$_{\rm eq}$--N$_{\rm H_2}$ correlation holds. Here, we  
    find inverse radial trends for  B$_{\rm eq}$ and N$_{\rm H_2}$ toward the center (Figure~\ref{fig:radial1}), which is probably caused by a feedback from  Sgr\,A$^{\star}$. The smooth increase of the ionized gas and the warm dust emission (21\,$\mu$m emission) toward the center is in favor of heating and ionization by  a feedback from  Sgr\,A$^{\star}$ (SMBH or a past AGN phase). 

    \item  Searching for a connection between the magnetic field and SFR in the CND, we find that the slope of the relationship between the magnetic field and the 70\,$\mu$m emission shows better agreement with theoretical predictions than with the 21\,$\mu$m emission. The  B$_{\rm eq}$--70\,$\mu$m  correlation follows the relation expected as if the magnetic field is amplified by small-scale turbulent dynamos in star-forming regions.  
    This also implies that the 70\,$\mu$m emission is primarily  {\FT due to dust heated by} young massive stars, while the 21\,$\mu$m emission {\apjrv is} contaminated by radiation from different sources.

    \item Computing the energy content of the ISM that encompasses the thermal, nonthermal, and kinetic energy density, we discover that the nonthermal (magnetic field and cosmic rays) and kinetic energy densities dominate the thermal energy density. 
    The ISM of this central region is occupied by a low-$\beta$ plasma ($\rm E_{th}<<E_{b}$) and the turbulent motions are supersonic ($\rm E_{th}<<E_k$). {\FT We show that in addition to turbulence, cosmic rays and magnetic fields insert significant nonthermal pressures 
    in the circumnuclaer region.}

    \item The intrinsic mass-to-magnetic flux ratio mapped in the circumnuclear region shows that the ISM is subcritical ($\mu_{int} \lesssim \mu_{crit}\sim1-2$), indicating that the magnetic field can protect molecular clouds against gravitational collapse.

\end{itemize}

In conclusion, our findings suggest that the ISM surrounding the Sgr\,A$^{\star}$ exhibits complex conditions. Even under the equipartition assumption, underestimating the field strength, the magnetic fields play a crucial role in the formation of clouds and stars in this region. Similar studies at higher resolutions with the upcoming SKAO observations will uncover the mysteries of the presence of young massive stars in this region.

\begin{acknowledgments}
We would like to thank the referee for their valuable feedback, which was instrumental in improving the paper. Our thanks also extend to Prof. Farhad Yusef-Zadeh and Prof. Rainer Beck for their precious and constructive comments, which greatly contributed to enhancing the quality of this paper. This paper makes use of the following ALMA data: ADS/JAO.ALMA$\#$2021.1.00172. ALMA is a partnership of ESO (representing its member states), NSF (USA) and NINS (Japan), together with NRC (Canada), NSTC and ASIAA (Taiwan), and KASI (Republic of Korea), in cooperation with the Republic of Chile. 

F. M acknowledges financial support from the School of Astronomy at the Institute for Research in Fundamental Sciences-IPM. L.C. acknowledges support from grant no. PID2022-136814NB-I00 by the Spanish Ministry of Science, Innovation and Universities/State Agency of Research MICIU/AEI/10.13039/501100011033 and by ERDF,UE. V.M.R. acknowledges support from the grant PID2022-136814NB-I00 by the Spanish Ministry of Science, Innovation and Universities/State Agency of Research MICIU/AEI/10.13039/501100011033 and by ERDF, UE;  the grant RYC2020-029387-I funded by MICIU/AEI/10.13039/501100011033 and by "ESF, Investing in your future", and from the Consejo Superior de Investigaciones Cient{\'i}ficas (CSIC) and the Centro de Astrobiolog{\'i}a (CAB) through the project 20225AT015 (Proyectos intramurales especiales del CSIC); and from the grant CNS2023-144464 funded by MICIU/AEI/10.13039/501100011033 and by “European Union NextGenerationEU/PRTR”. A.S.-M.\ acknowledges support from the RyC2021-032892-I grant funded by MCIN/AEI/10.13039/501100011033 and by the European Union `Next GenerationEU’/PRTR, as well as the program Unidad de Excelencia María de Maeztu CEX2020-001058-M, and support from the PID2023-146675NB-I00 (MCI-AEI-FEDER, UE). S.Z.\ acknowledges support from the Strategic Priority Research Program of the Chinese Academy of Sciences (CAS) Grant No.\ XDB0800303.
\end{acknowledgments}

%




\appendix

\section{Parameters in Equipartition Magnetic Field Equation}\label{sec:appendix b}
In the revised equipartition magnetic field strength equation presented by \cite{beck2005revised} (Equation~(\ref{eq:mf})), the complete expressions of $C_1$ and $C_2$ are as follows

\begin{center}
$C_1 = 3e/(4\pi m_e^3c^5) = 6.26428\times 10^{18}\,\rm erg^{-2}s^{-1}G^{-1}$
\end{center}

\begin{center}
$C_2(\alpha_n) = \frac{1}{4}C_3(\alpha_n+\frac{5}{3})/(\alpha_n+1)\Gamma[(3\alpha_n+1)/6]\times\Gamma[(3\alpha_n+5)/6],$
\end{center}

where the $\Gamma$ is the numerical gamma function and $\alpha_n$ and $i$ are nonthermal spectral index and inclination. In the equation above, $C_3$ is

\begin{center}
$C_3 = \sqrt{3}e^3/(2\pi m_eC^2)= 1.86558\times10^{-23}\rm erg G^{-1}sr^{-1}$,
\end{center}

For the regular and constant inclination concerning the sky plane, $\rm C_4 = [cos(i)]^{\alpha_n+1}$ and for the case of our studied region, the inclination is considered to be $72^{\circ}$ \citep{ferriere2007spatial,sawada2004molecular}. Also, $\alpha_n$ is assumed to be 0.7 for the pure synchrotron cooling of CREs \citep{hsieh2017molecular,goicoechea2018high}.



\bibliography{reference}{}
\bibliographystyle{aasjournal}



\end{document}